%% file: main.tex
\documentclass[twocolumn,english,aps,amsmath,amssymb,reprint,nofootinbib,floatfix,showkeys,superscriptaddress]{revtex4-2}

\usepackage{lineno} 
\usepackage{appendix}

\usepackage[table]{xcolor}
\usepackage{graphicx}
\usepackage{dcolumn}
\usepackage{bm}
\usepackage[colorlinks = true,citecolor=blue,linkcolor = blue,urlcolor=blue]{hyperref}

\usepackage[capitalise]{cleveref}
\usepackage{graphicx}	
\usepackage{amsmath}	
\usepackage{amssymb}	
\usepackage{wasysym}    
\usepackage{mathtools}  
\usepackage{aas_macros}
\usepackage{array,multirow,graphicx}  
\usepackage{multirow}

\begin{document}

\title{A measurement of $H_0$ from DESI DR1 using energy densities}







\author{Alex Krolewski}
\email{akrolews@uwaterloo.ca}
\affiliation{Waterloo Centre for Astrophysics, University of Waterloo, Waterloo, ON N2L 3G1, Canada}
\affiliation{Department of Physics and Astronomy, University of Waterloo, Waterloo, ON N2L 3G1, Canada}

\author{Andrea Crespi}
\author{Will J. Percival}
\affiliation{Waterloo Centre for Astrophysics, University of Waterloo, Waterloo, ON N2L 3G1, Canada}
\affiliation{Department of Physics and Astronomy, University of Waterloo, Waterloo, ON N2L 3G1, Canada}
\affiliation{Perimeter Institute for Theoretical Physics, 31 Caroline St. North, Waterloo, ON NL2 2Y5, Canada \\
The affiliations of the remaining authors are listed in Appendix D.}

\author{Marco Bonici}%
\affiliation{Waterloo Centre for Astrophysics, University of Waterloo, Waterloo, ON N2L 3G1, Canada}
\affiliation{Department of Physics and Astronomy, University of Waterloo, Waterloo, ON N2L 3G1, Canada}

\author{Hanyu Zhang}
\affiliation{Waterloo Centre for Astrophysics, University of Waterloo, Waterloo, ON N2L 3G1, Canada}
\affiliation{Department of Physics and Astronomy, University of Waterloo, Waterloo, ON N2L 3G1, Canada}




\author{J. Aguilar\textsuperscript{4}, S. Ahlen\textsuperscript{5}, D. Bianchi\textsuperscript{6, 7}, D. Brooks\textsuperscript{8}, R. Canning\textsuperscript{9}, E. Chaussidon\textsuperscript{4}, T. Claybaugh\textsuperscript{4}, A. Cuceu\textsuperscript{4}, S. Cole\textsuperscript{10}, A. de la Macorra\textsuperscript{11}, J. Della Costa\textsuperscript{12,13},
P. Doel\textsuperscript{8}, J. Edelstein\textsuperscript{14, 15},
S. Ferraro\textsuperscript{4, 15}, A. Font-Ribera\textsuperscript{16}, J. Forero-Romero\textsuperscript{17, 18}, E. Gaztañaga\textsuperscript{19, 9, 20}, S. Gontcho a Gontcho\textsuperscript{4, 21},
G. Gutierrez\textsuperscript{22}, J. Guy\textsuperscript{4}, H. Herrera-Alcantar\textsuperscript{23,24}, K. Honscheid\textsuperscript{25, 26, 27}, D. Huterer\textsuperscript{28, 29}, M. Ishak\textsuperscript{30}, D. Joyce\textsuperscript{13}, R. Kehoe\textsuperscript{31}, D. Kirkby\textsuperscript{32}, T. Kisner\textsuperscript{4}, A. Kremin\textsuperscript{4}, O. Lahav\textsuperscript{8}, C. Lamman\textsuperscript{27}, M. Landriau\textsuperscript{4}, L. Le Guillou\textsuperscript{33}, M. Levi\textsuperscript{4}, M. Manera\textsuperscript{34, 16}, A. Meisner\textsuperscript{13}, R. Miquel\textsuperscript{35, 16}, J. Moustakas\textsuperscript{36},
A. Muñoz-Gutiérrez\textsuperscript{11},
S. Nadathur\textsuperscript{9}, 
G. Niz\textsuperscript{37, 38}, N. Palanque-Delabrouille\textsuperscript{24, 4}, C. Poppett\textsuperscript{4, 14, 15}, F. Prada\textsuperscript{39}, I. Pérez-Ràfols\textsuperscript{40}, G. Rossi\textsuperscript{41}, L. Samushia\textsuperscript{42, 43, 44}, E. Sanchez\textsuperscript{45}, D. Schlegel\textsuperscript{4}, M. Schubnell\textsuperscript{28, 29},  J. H. Silber\textsuperscript{4}, D. Sprayberry\textsuperscript{13}, G. Tarlé\textsuperscript{29}, B. A. Weaver\textsuperscript{13}, and H. Zou\textsuperscript{46}}


\begin{abstract}
We present a new measurement of the Hubble constant, independent of standard rulers and robust to pre-recombination modifications such as Early Dark Energy (EDE), obtained by calibrating the total energy density of the Universe. We start using the present-day photon density as an anchor, and use the baryon-to-photon ratio from Big Bang Nucleosynthesis based measurements and the baryon-to-matter ratio from the baryons' imprint on galaxy clustering to translate to a physical matter density at present day. We then compare this to measurements of the ratio of the matter density to the critical density ($\Omega_{\mathrm{m}}$), calculated using the relative positions of the baryon acoustic oscillations, to measure the critical density of the universe and hence $H_0$. The important measurements of the evolution of the energy density all happen at low redshift, so we consider this a low-redshift measurement. We validate our method both on a suite of $N$-body mocks and on noiseless theory vectors generated across a wide range of Hubble parameters in both $\Lambda$CDM and EDE cosmologies. Using DESI DR1 data combined with the angular CMB acoustic scale and the latest BBN constraints, we find $H_0 = 69.0 \pm 2.5$ km s$^{-1}$ Mpc$^{-1}$, consistent with existing early and late-time determinations of the Hubble constant. We consider the impact of non-standard dark energy evolution on our measurement. Future data, including that from further iterations of DESI and from Euclid, will add to these results providing a powerful test of the Hubble tension.
\end{abstract}

\maketitle

\section{Introduction}


The $\Lambda$CDM model has been incredibly successful at explaining cosmic microwave background and large-scale structure measurements, with its parameters constrained to very high precision \cite{Planck:2020,DESI2024.VI.KP7A,DESI2024.VII.KP7B}.
However, a potential crack has emerged with the high-significance tension between local measurements of the Hubble constant using the local distance ladder from Cepheids calibrating type Ia supernovae \cite{Riess-Hubble}, $H_0 = 73.04 \pm 1.04$ km s$^{-1}$ Mpc$^{-1}$, and cosmological inference in the $\Lambda$CDM model from Planck CMB measurements, $H_0 = 67.37 \pm 0.54$ km s$^{-1}$ Mpc$^{-1}$. Large-scale structure measurements from the baryon acoustic oscillation (BAO) standard ruler favor a similarly low $H_0$ measurement (when calibrated by BBN), with $H_0 = 68.53 \pm 0.80$ km s$^{-1}$ Mpc$^{-1}$ from DESI DR1 \cite{DESI2024.III.KP4} and $68.51 \pm 0.58$ km s$^{-1}$ Mpc$^{-1}$  from DESI DR2 \cite{DESIDR2BAO}. This difference could be due to new physics \cite{DiValentino21}
or systematics in one or more datasets \cite{Efstathiou20}.

A number of alternative $H_0$ measurements have been made with different distance ladder calibrators or other astrophysical objects entirely. Using the Tip of the Red Giant Branch and J-Region Asymptotic Giant Branch stars in addition to Cepheids, \cite{Freedman24} find $H_0 = 70.0 \pm 1.54$ km s$^{-1}$ Mpc$^{-1}$ (adding their statistical and systematic errors in quadrature).
On the other hand, \cite{Riess24} find a slightly higher value of $H_0$ from these same three calibrators in JWST data, $H_0 = 72.6 \pm 2.0$ km s$^{-1}$ Mpc$^{-1}$. Other alternative methods are promising for the future but currently less constraining.
For example, lensing time delays are statistically powerful but systematically limited by degeneracies with the mass profile \cite{Birrer20}.
Constraining $H_0$ from the single standard siren GW170817 achieves a $\sim$15\% precision \cite{Abbott17}, with considerable improvements possible by using outflow emission to constrain its inclination angle \cite{Hotokezaka19,Wang23,Palmese23}. This method could be quite powerful in the future if a larger sample of gravitational wave events with confirmed counterparts can be discovered.

A popular class of new physics resolutions to the Hubble tension involves models that change the sound horizon $r_d$ \cite{KnoxMillea20}.
One such model involves adding a new component of ``early dark energy'' modifying the expansion rate around recombination \cite{Karwal:2016,Polin:2019} (see also its relatives New Early Dark Energy \cite{Niedermann20} and Early Modified Gravity \cite{Braglia20}). Other possibilities include modifications to the recombination history \cite{Lynch24,Mirpoorian24}, which could arise from variations in fundamental constants \cite{Chluba23}, from primordial magnetic fields \cite{Jedamzik20,GarciaEscudero22},
or from variations in the electron mass \cite{SchoenebergVacher24} (see also \cite{H0Olympics} for a comprehensive comparison of various theoretical solutions to the Hubble tension).

These models have motivated cosmological $H_0$ constraints from large-scale structure independent of the sound horizon, by marginalizing over an additional parameter scaling $r_d$ \cite{BaxterSherwin21,Philcox21c,Philcox22,Farren22,Madhavacheril24}. Recent sound horizon-free measurements from CMB lensing cross-correlations \cite{Farren24} and the DESI DR1 power spectrum \cite{Zaborowski24} have found $H_0$ consistent with other cosmological measurements, $64.3^{+2.1}_{-2.4}$ and $67.9^{+1.9}_{-2.1}$ km s$^{-1}$ Mpc$^{-1}$ respectively.\footnote{We quote the constraints of \cite{Zaborowski24} using $\Omega_{\mathrm{m}}$ constrainted by Pantheon+ supernovae; using Union3 supernovae gives a very similar $H_0$ measurement while using DESY5 supernovae shifts $H_0$ by 1.2 km s$^{-1}$ Mpc$^{-1}$.}
Combining these constraints, and replacing $\Omega_\mathrm{m}$ information from supernovae with DESI DR1 BAO and the angular acoustic scale from the CMB, yields $H_0 = 69.2^{+1.3}_{-1.4}$ km s$^{-1}$ Mpc$^{-1}$ \cite{Zaborowski25}.
A similar method from \cite{Brieden23}, using uncalibrated BAO plus shape information from the model-agnostic ShapeFit method \cite{Brieden21,Brieden22},
yields $H_0 = 70.1^{+1.9}_{-2.1}$ from the BOSS and eBOSS galaxy power spectrum.
Finally, \cite{Kalus23} measured the power spectrum turnover scale in the BOSS quasar power spectrum and updated to the DESI power spectrum in \cite{Kalus25}. Since they only use the turnover scale itself and not any shape information on smaller scales, their $H_0$ constraint is weaker, $74.0^{+7.2}_{-3.5}$ km s$^{-1}$ Mpc$^{-1}$.

These constraints generally assume a $\Lambda$CDM model for the shape of the power spectrum, beyond the marginalization over $r_d$. If a Hubble-tension-resolving modification also changes the shape of the power spectrum (e.g.\ as shown for EDE in \cite{Hill20}), it could shift these sound horizon-independent constraints.
For instance, \cite{Smith22} find that the sound horizon marginalized constraints on $H_0$, when analyzed in an early dark energy cosmology,
shift towards higher values of $H_0$ by $\Delta H_0 = 3.8$ km s$^{-1}$ Mpc$^{-1}$ and the error nearly doubles. \cite{Kable24} also finds that sound horizon independent $H_0$ constraints can be consistent with a high value of the Hubble constant.
Likewise, the constraints of \cite{Brieden23} are also somewhat model dependent: allowing for a wide prior on $n_s$, freeing the neutrino mass, or adding $\Delta N_{\textrm{eff}}$ broadens their constraint to $\sigma_{H_0} = 2.5$ (3.2, 2.5) km s$^{-1}$ Mpc$^{-1}$, while leaving the central value largely unchanged except for $\nu\Lambda$CDM, which increases $H_0$ to 73.3 km s$^{-1}$ Mpc$^{-1}$.

In light of these concerns, an alternative $H_0$ measurement has been proposed that limits the processes modeled to the very robust physics driven by the background energy densities of the Universe \cite{BOSSBAOAmp,Krolewski24b,Crespi25}. Specifically, the measurement is anchored to the extremely precise measurement of the CMB photon density; it then uses BBN based measurements to constrain the baryon-to-photon ratio and galaxy clustering to constrain the baryon fraction $\Omega_{\mathrm{b}} / \Omega_{\mathrm{m}}$; together these give the physical matter density at present day. When comparing against $\Omega_{\mathrm{m}}$ from the low redshift geometry of the Universe (e.g.\ BAO or supernovae), this allows us to infer the critical energy density of the Universe, which directly gives $H_0$. In \cite{BOSSBAOAmp}, this method was explicitly shown to recover unbiased measurements of $H_0$ in EDE cosmologies with different true values of $H_0$, showing that it is unbiased by any broadband shifts in the galaxy power spectrum due to the early-time new physics.

This method has previously been applied to the BOSS data, finding $H_0 = 67.1^{+6.3}_{-5.3}$ km s$^{-1}$ Mpc$^{-1}$, limited by the precision of the baryon fraction \cite{BOSSBAOAmp,Krolewski24b}.
In this work, we improve the constraining power on the baryon fraction (and thus $H_0$) using data from DESI DR1. We also use a new, more robust pipeline to self-consistently model the baryon fraction, described in our companion paper \cite{Crespi25}.

In Section~\ref{sec:methods}, we describe the data used in this work and summarize the baryon fraction fitting pipeline. 
We test our pipeline on mocks in Section~\ref{sec:full_shape_mock_tests} for our fiducial
pipeline using Effective Field Theory fits to the power spectrum, and in Section~\ref{sec:BAO_mock_tests} using an alternative pipeline fitting to only the BAO in the post-reconstruction correlation function. 
We test on both the same $N$-body mocks used to validate the DESI full-shape measurements, and on a wide range of noiseless theory vectors in $\Lambda$CDM and EDE cosmologies.
Finally, in Section~\ref{sec:data}, we describe our cosmological constraints from the DESI DR1 data.
We did not run our pipeline on data until all of the mock tests in Section~\ref{sec:full_shape_mock_tests} and~\ref{sec:BAO_mock_tests} were satisfied. 

Throughout, all power spectra and correlation functions are measured using the DESI fiducial cosmology,
with $\Omega_{\mathrm{b}} h^2 = 0.02237$, $\Omega_c h^2 = 0.1200$, $h = 0.6736$, $A_s = 2.083 \times 10^{-9}$, $n_s = 0.9649$, and a single massive neutrino with mass 0.06 eV.

\section{Methods}
\label{sec:methods}

\subsection{Measuring $H_0$ from energy densities}

Our method to measure $H_0$ is straightforward: we estimate
the critical energy density of the Universe at present day, and therefore $H_0$. The method can be summarized with this relation:
\begin{equation}  
    \frac{3c^2H_0^2}{8\pi G} = \epsilon_c=
    \underbrace{\vphantom{
\frac{\epsilon_{b,0}}{\epsilon_{\gamma,0}} 
    }
    \epsilon_{\gamma,0} }_{\textrm{(i)}}
    \times
    \underbrace{
    \frac{\epsilon_{b,0}}{\epsilon_{\gamma,0}}}_{\textrm{(ii)}}
    \times
    \underbrace{\frac{\epsilon_{m,0}}{\epsilon_{b,0}}}_{\textrm{(iii)}}
    \times
    \underbrace{\frac{1}{\Omega_{m,0}}}_{\textrm{(iv)}}\,, 
\label{eq:basic_eqn}
\end{equation}
where $\epsilon$ refers to a background energy density. Methods to robustly measure (i), (ii) and (iv) from simple physical measurements have already been developed using existing data; (i) can be measured from the CMB temperature, (ii) from BBN constraints on the photon-baryon ratio, and (iv) from geometric measurements of $\Omega_{\mathrm{m}}$. It was recently shown that (iii) can be measured in a robust way from galaxy clustering \cite{BOSSBAOAmp}. This measurement relies on the fact that the gravitational growth of perturbations is driven by the summed potential of baryons and dark matter. As a result, the amplitude of baryon effects in the low-redshift matter power spectrum depends only on the baryon fraction.

In order to extract the baryon fraction from only the effect of the growth of perturbations, we adopt a new procedure \cite{Crespi25}. This procedure adds an additional parameter $\gamma_b$ that changes the relative weighting of the baryon and CDM components in the gravitational potential adopted by the Boltzmann solver \texttt{CAMB}, which models the evolution of perturbations in time. Although unphysical, when $\gamma_b$ is treated as a free parameter, this approach allows us to extract the baryon fraction from only the growth of perturbations in a fit to data. Performing a full fit with galaxy clustering data and extracting the constraint on the cosmological parameters $\Omega_{\mathrm{b}}/\Omega_{\mathrm{m}}$ would bring in additional information on $\Omega_{\mathrm{b}}$ (such as from the positions of the BAO peaks), complicating the interpretation as multiple physical processes contribute to the measurement.

We are able to constrain $\gamma_b$ in a fit because the growth of baryonic perturbations is different from the growth of CDM perturbations, leading to features in the combined matter power spectrum, the amplitude of which depend on $\gamma_b$. In particular, the baryon perturbations can't grow before decoupling due to radiation pressure from the photons, but instead oscillate with the photons leading to a series of peaks and troughs in the power spectrum of the baryons perturbations as a function of scale, called Baryon Acoustic Oscillations (BAO). In addition to the oscillating signal, since baryon perturbations don't grow before decoupling, a larger baryon fraction suppresses the power spectrum on scales smaller than the sound horizon. On the other hand, the CDM perturbations begin to grow earlier, after matter-radiation equality. 

To obtain this information, we previously \cite{BOSSBAOAmp} approximated these effects by splitting the linear power spectrum into baryon and CDM transfer functions. We then allowed $\gamma_b$ to re-weight the sum of the transfer functions. However, this approach relies on ambiguous definitions of the baryon and CDM transfer functions. In this work, when constraining the baryon fraction from the full shape of the power spectrum (using the Effective Field Theory of large scale structure), we use the new approach of \cite{Crespi25}. In this approach, the new parameter $\gamma_b$ directly and self-consistently affects the growth of perturbations in the Boltzmann solver \texttt{CAMB}. 
We replace the cosmological baryon fraction (determined from $\Omega_\mathrm{b}$ and $\Omega_\mathrm{m}$) with our new parameter $\gamma_b$ in all equations governing structure growth, while retaining the cosmological baryon fraction in the background and thermal history of the Universe.



\subsection{Data used in our analyses}

The Dark Energy Spectroscopic Instrument (DESI) \cite{DESI2016a.Science,DESI2016b.Instr} is conducting an eight-year survey across 17,000 deg$^2$ \cite{SurveyOps.Schlafly.2023}, with the goal of updated goal of spectroscopically confirming 63M galaxies. To obtain these spectra, 5,000 fibers \cite{FiberSystem.Poppett.2024} are robotically positioned across a 7 deg$^2$ field \cite{FocalPlane.Silber.2023,Corrector.Miller.2023} using the Mayall 4-m telescope at Kitt Peak National Observatory \cite{DESI2022.KP1.Instr}. Survey operations and assignment to DESI tiles are described in \cite{SurveyOps.Schlafly.2023}.
Once observed, spectra are processed with an automated spectroscopic reduction pipeline \cite{Spectro.Pipeline.Guy.2023}, and redshifts are determined automatically for all targets \cite{RedrockQSO.Brodzeller.2023}.

In this work, we use catalogs and clustering measurements from the first year of DESI observations, DESI DR1 \cite{DESI2024.I.DR1,KP3s15-Ross}. We use measurements of the pre-reconstruction power spectrum and the post-reconstruction correlation function. In both cases, we use the two-point functions calculated for analyses of the full-shape \cite{DESI2024.V.KP5} and BAO position \cite{DESI2024.III.KP4} signals. These papers contain more details about the analysis choices, which we summarize here. Clustering measurements were calculated using weights on each galaxy that mitigated the impact of imaging systematics, redshift failures, and incompleteness due to fiber collisions (see \cite{KP3s15-Ross} for a comprehensive description of these weights), and optimse the signal. The power spectrum multipoles are fit in the $k$ range [0.02, 0.20] $h$ Mpc$^{-1}$. Fiber collisions are mitigated by the `$\theta$-cut' method, which removes all pairs separated by angular scales $<0.05^\circ$ \cite{KP3s5-Pinon}.

When we fit to the post-reconstruction correlation function, we jointly extract the baryon fraction and BAO positions required to constrain $\Omega_{\mathrm{m}}$. When we fit the power spectrum, we combine with pre-computed Alcock-Paczynski (AP) parameters from post-reconstruction BAO to determine $\Omega_{\mathrm{m}}$, using a joint covariance estimated from 1000 EZMock \cite{EZMocks} power spectrum and AP parameter measurements. An empirical scaling factor was applied to the covariance matrix, to account for inaccuracies in the EZMocks and approximate fiber assignment \cite{DESI2024.V.KP5}. A systematic error covariance accounts for imaging systematics, fiber assignment, and HOD-dependent systematics and prior-weight effects \cite{KP5s7-Findlay}. For our fiducial analysis in which we measure the BAO amplitude from the full-shape of the galaxy power spectrum, we use the LRG3 BAO measurement at $0.8 < z < 1.1$ rather than the ELG+LRG measurement used in \cite{DESI2024.III.KP4} for ease of calculating the covariance between the post-reconstruction correlation function and the pre-reconstruction power spectrum. We also include the Ly$\alpha$ forest BAO measurement at $z=2.33$ \cite{DESI2024.IV.KP6}. 

When fitting $\gamma_b$ to the post-reconstruction correlation function, we use a broader scale range than the DESI analysis \cite{DESI2024.III.KP4} ($28 < r < 180$ $h^{-1}$ Mpc), since our pipeline includes an additional broadband term that is better constrained with the broader scale range. For these fits, we do not include a correction for fiber assignment, since the BAO scale is well-separated from the small scales impacted by fiber collisions. Covariances were computed analytically \cite{Rashkovetskyi23}.

The analysis pipelines were validated using mock data from the Abacus-2 mocks (described in Section 3.1 of \cite{DESI.DR2.BAO.cosmo}) These were produced from the large suite of high-resolution $N$-body AbacusSummit simulations \cite{Garrison19,Maksimova21}. Galaxies, with HODs chosen to match DESI, were populated into the 25 base boxes, generated with a Planck $\Lambda$CDM cosmology in a (2 $h^{-1}$ Gpc)$^3$ box. These mocks were processed through the DESI survey geometry, and were generated with realistic runs of the fiber assignment process (``altmtl'') \cite{KP3s7-Lasker}.

We use the $\theta_\star$ measurement from Planck PR3, $100 \theta_\star = 104.089 \pm 0.031$ (the ``combined'' constraint from \cite{Planck:2020}).
We also include a constraint on the physical baryon density from Big Bang Nucleosynthesis (BBN) \cite{Schoneberg24}, $\Omega_{\mathrm{b}} h^2 = 0.02218 \pm 0.00055$. We choose the BBN measurement of $\Omega_{\mathrm{b}} h^2$ to be as independent of the CMB as possible. This measurement relies on primordial deuterium and helium measurements (as compiled by the Particle Data Group \cite{PDG}) and nuclear rates from the \texttt{PRyMordial} code \cite{Burns24}. This measurement is quite insensitive to additional relativistic species $\Delta N_{\textrm{eff}}$, and uncertainties in the reaction rates as well as systematic errors in the abundances are already marginalized over.

\subsection{Energy density $H_0$ in EFT fits to the power spectrum}

We constrain $H_0$ using Equation~\ref{eq:basic_eqn} by adding new ``density'' parameters $h^\textrm{dens}$, $\omega_{\mathrm{b}}^{\textrm{dens}}$ and $\Omega_{\mathrm{m}}^{\textrm{dens}}$ to our analysis. These are used in the analysis in such a way that they only depend on the physical processes selected. By treating the standard cosmological parameters $H_0$, $\Omega_{\mathrm{m}}$, $\Omega_{\mathrm{b}} h^2$, $n_s$ and $A_s$ as nuisance parameters to be marginalized over, we extract only the information required by Equation~\ref{eq:basic_eqn} using the desired physical processes, but can still perform a joint fit. 

BBN information on $\omega_b^{\textrm{dens}}$ is implemented as a Gaussian prior. As a result, we will often refer to our analysis as including only two new parameters ($h^\textrm{dens}$ and $\Omega_{\mathrm{m}}^{\textrm{dens}}$) and won't show $\omega_b^{\textrm{dens}}$ contours in our tests on mocks.

From $h^\textrm{dens}$ and $\omega_{\mathrm{b}}^{\textrm{dens}}$ we define $\gamma_b$:
\begin{equation}
\gamma_b = \frac{\omega_{\mathrm{b}}^{\rm dens}}{{\Omega_{\mathrm{m}}^{\rm dens} (h^{\rm dens})^2} - \omega_{\mathrm{massive}-\nu}}
\label{eqn:gammaB_definition}
\end{equation}
where $\omega_{\mathrm{massive}-\nu}$ is the density of massive neutrinos, for which we fix the neutrino mass at 0.06 eV. We then insert $\gamma_b$ into our modified version of \textsc{CAMB} and use the resulting linear power spectrum as the input for the perturbative model of galaxy clustering.

The geometric information we include is the relative BAO positions, compressed to the AP parameters $\alpha_{\textrm{iso}}$ and $\alpha_{\textrm{AP}}$, and fitted to the post-reconstruction correlation function. We isolate the anisotropic information by marginalizing over the isotropic BAO position using an extra free parameter to model these measurements only.
The model values of $H(z)$ and $D_A(z)$ required to fit these data are calculated using $\Omega_{\mathrm{m}}^{\rm dens}$.

\begin{figure*}
    \includegraphics[width=\textwidth]{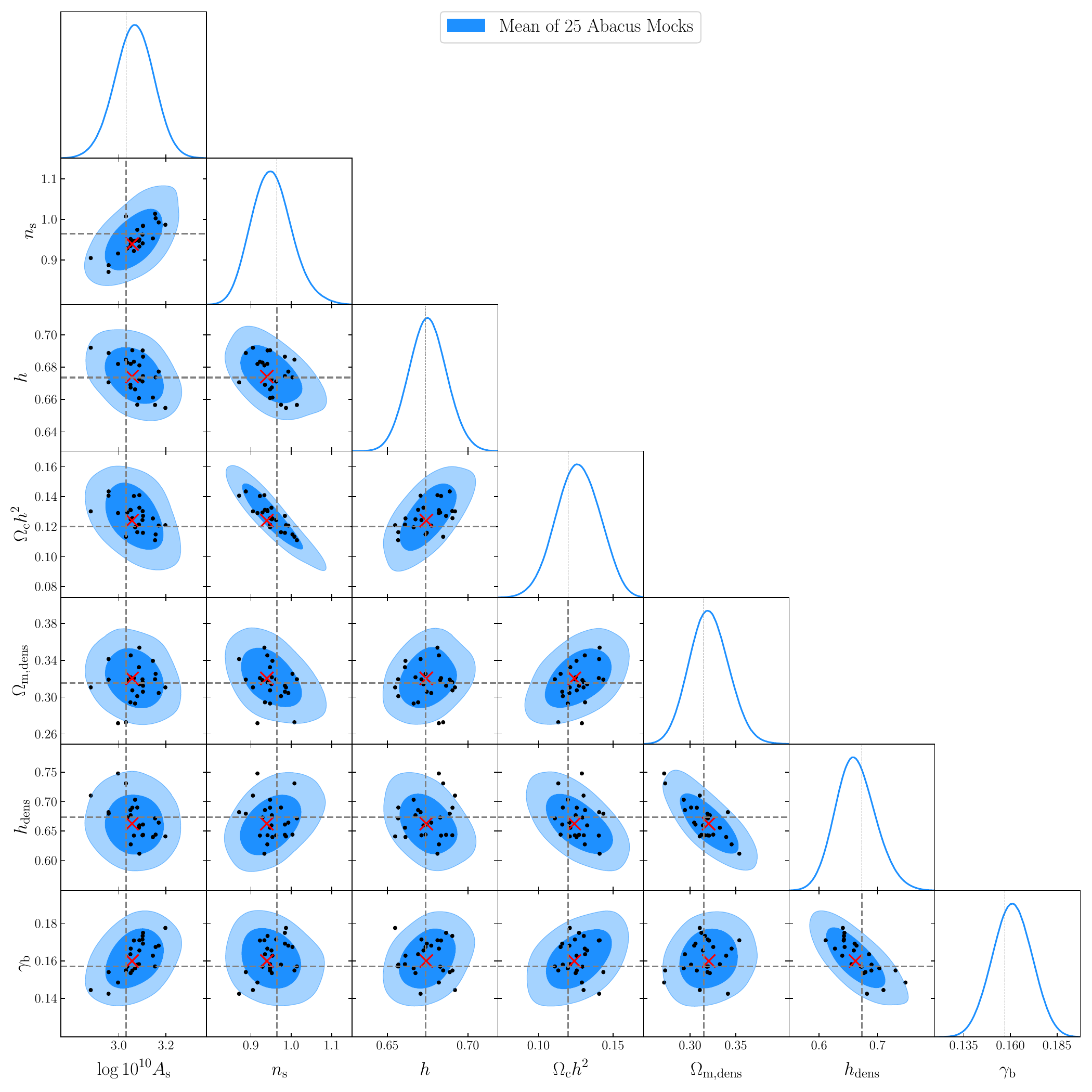}
    \caption{Constraints on $\Lambda$CDM parameters and the Hubble parameter $h^{\textrm{dens}}$ derived from the combination of energy densities, using the mean of the 25 Abacus mocks (truth in dashed lines). The red crosses show the maximum of the posterior. Black dots show marginalized means from the individual mocks.}
    \label{fig:abacus_mock_test}
\end{figure*}

We use the Effective Field Theory of large scale structure (EFT of LSS) to model the galaxy power spectrum monopole and quadrupole.  Specifically, we model the galaxy power spectrum using the Eulerian EFT formalism implemented in 
\textsc{velocileptors} \cite{Chen20,Chen21} and emulated by \textsc{Effort.jl} \cite{Effort}. 
The galaxy power spectrum model uses a symmetry-based bias expansion with linear and second-order bias parameters, counterterms and stochastic parameters to marginalize over unknown small-scale physics. We do not consider relative velocity biases \citep[e.g.][]{Schmidt16,Chen19a,KP4s2-Chen} in our model. These biases can cause offsets in the BAO peak positions and thus potentially affect inference of the baryon fraction. Their effect is expected to be small compared to our current precision on the baryon fraction, but may need to be marginalized over in the future. An approximate estimate of their effect is given in Appendix~\ref{sec:relative_velocity}, and full implementation is left for future work.

The
\textsc{velocileptors} Eulerian EFT model
uses 1-loop resummed Eulerian perturbation theory (REPT) where the linear and next-to-leading order terms are split into wiggle and no-wiggle components (following the method of \cite{Wallisch18}) and the wiggle part is exponentially damped with some dispersion $\Sigma$ that is a function of the linear power spectrum. This is described in detail in Appendix A of \cite{Chen20}.
While in principle the damping $\Sigma$ may be degenerate with $\gamma_b$, we found this was not the case when
fitting to the BOSS pre-reconstruction correlation function
in \cite{BOSSBAOAmp}; we additionally found that BAO reconstruction reduces $\Sigma$ but does not appreciably improve the constraining power on $\gamma_b$.

We use 
the same model as the one used in the DESI DR1 full-shape analysis \cite{DESI2024.V.KP5}. In our previous work \cite{BOSSBAOAmp}, we used \texttt{CLASS-PT} instead, a different code also using Eulerian EFT to model the galaxy power spectrum. In Appendix~\ref{sec:compare_to_old_method}, we extensively compare to the old pipeline and show that it recovers the same results.

In a Bayesian analysis with a large number of poorly constrained nuisance parameters entering the likelihood nonlinearly, prior volume projection effects \cite{Hamann12,Simon23,Carrilho23} are a generic issue. Projection effects are shifts in the marginalized likelihood relative to the posterior maximum (MAP). They can cause biases in marginalized posteriors when testing on mock datasets, even if the MAP is unbiased. They were extensively studied in the context of the DESI DR1 $\Lambda$CDM and $w_0 w_a$CDM full-shape constraints in \cite{DESI2024.V.KP5,KP5s1-Maus}. In our analysis, adding two extra cosmological parameters ($\Omega_{\mathrm{m}}^{\textrm{dens}}$ and $h^{\textrm{dens}}$) makes the projection effects even more severe.

We therefore impose Halo Occupation Distribution (HOD) informed priors \cite{Zhang24,Ivanov24a,Ivanov24b,Zhang25} on the bias, counterterm, and stochastic parameters, allowing us to nearly eliminate projection effects (as shown in Sec.~\ref{sec:full_shape_mock_tests} below). This method maps broad priors on HOD models of galaxy formation to EFT nuisance parameters, reducing the volume of unphysical nuisance parameters and thereby markedly reducing projection effects \cite{Zhang24}.
We follow the implementation of \cite{Zhang24,Zhang25}, who 
trained a normalizing flow model to learn the mapping from broad HOD priors to priors on the EFT nuisance parameters. 
In \cite{Crespi25}, we show that the HOD prior recovers unbiased cosmology when fitting noiseless theory vectors where the growth and cosmological baryon fractions differ (even though the HOD prior was trained on standard cosmology where these two coincide).


Because we use \textsc{Effort.jl}, we can
take advantage of other high-performance, state-of-the-art tools in \textsc{Julia} to dramatically decrease the runtime of our analysis. The likelihood built from \textsc{Effort.jl} is automatically differentiable.
We sample using the No U-Turn Sampler (NUTS) \cite{NUTS}.

\subsection{Energy density $H_0$ in post-reconstruction BAO fits}

Although we adopt the full clustering fits as our baseline, we compare against results fitting just the post-reconstruction BAO signal as a test of the robustness of our method. For these we use the method of \cite{Crespi25} to create linear power spectra including the $\gamma_b$ parameter to control the baryon fraction in gravitational perturbation growth, while keeping the background and thermal history fixed to that specified by $\Omega_\mathrm{b}$ and $\Omega_\mathrm{m}$.

We use this linear matter power spectrum as input to the extraction of BAO positions from the post-reconstruction correlation function. 
We largely follow the standard DESI method for BAO fitting \cite{KP4s2-Chen,DESI2024.III.KP4}, though with a slightly different model for the no-wiggle power spectrum.
This splits the linear power spectrum into wiggle ($P_{\textrm{w}}(k)$) and no-wiggle ($P_{\textrm{nw}}(k)$) parts. The post-reconstruction correlation function  is then modelled with a damped linear power spectrum
\begin{equation}
    P(k,\mu) = \mathcal{B}(k,\mu) P_{\textrm{nw}}(k) + \mathcal{C}(k,\mu)P_{\textrm{w}}(k) + \mathcal{D}(k,\mu)
\end{equation}
where $\mathcal{B}(k,\mu)$ matches \cite{DESI2024.III.KP4} with an additional term proportional to $b_\partial k^2$ \cite{Ding18}:
\begin{align}
    \mathcal{B}(k,\mu) = \left((b_1 + f\mu^2)^2 + b_{\partial} \frac{k^2}{k_L^2} (b + f\mu^2)\right) \nonumber \\ \times
    \left(\frac{1}{1 + 0.5 k^2 \mu^2 \Sigma_s^2}\right)^2
\label{eqn:4}
\end{align}
and $\mathcal{C}(k,\mu)$ is exactly the same as in \cite{DESI2024.III.KP4}
\begin{equation}
    \mathcal{C}(k,\mu) = (b_1 + f\mu^2)^2 \exp\left[-\frac{1}{2}k^2 \left(\mu^2 \Sigma_{\parallel}^2 + (1 - \mu^2) \Sigma_\perp^2\right)\right]
\end{equation}
The broadband term $\mathcal{D}(k,\mu)$ is identical to the spline basis used in \cite{DESI2024.III.KP4}:
\begin{equation}
\mathcal{D}_\ell(k) = \sum_{n=-1}^{n_\textrm{max}} a_{\ell,n} W_3 \left(\frac{k}{\Delta - n} \right)
\end{equation}
where $W_3$ is a piecewise cubic spline kernel, $\Delta = 2 \pi / r_d = 0.06$ $h$ Mpc$^{-1}$, and $n_{\textrm{max}} = 7$. This is an update on the method used in \cite{BOSSBAOAmp}, which adopted polynomial broadband terms, which we found to be quite degenerate with the BAO amplitude, particularly the $1/s^2$ term. 
In contrast, the cubic spline basis is not degenerate with the BAO amplitude by construction, as it marginalizes over any (Fourier-space) oscillatory component with a higher frequency than the BAO as well as large scale systematics from Fourier modes with $k < k_{\textrm{min}}$.
Therefore, we use the same cubic spline broadband parameters as \cite{DESI2024.III.KP4}, with two terms affecting the monopole (a constant and $s^2$ term) and four terms affecting the quadrupole (constant, $s^2$, and Hankel transforms of the two lowest-order spline terms). We apply the same priors as \cite{DESI2024.III.KP4}, a flat infinite prior on the constant and $s^2$ terms and $\mathcal{N}(0,10^4)$ on the spline parameters.


Another difference from our previous analysis \cite{BOSSBAOAmp} is the treatment of the Finger-of-God small-scale redshift space distortion (RSD) term (second line in Eq.~\ref{eqn:4}), which we apply to the no-wiggle power spectrum only, following \cite{KP4s2-Chen}.
We also replace the narrow Gaussian prior on $b_\partial$ from \cite{BOSSBAOAmp} ($15 \pm 5$) with a broad flat prior between -1000 and 1000, increasing the fitted range to $28 < r < 180$ $h^{-1}$ Mpc to provide the necessary extra constraining power on $b_\partial$.
Finally, we use the ``RecSym'' reconstruction convention \cite{White15,KP4s3-Chen} with $\Sigma_{\textrm{sm}} = 15$ $h^{-1}$ Mpc for all tracers except QSO, for which $\Sigma_{\textrm{sm}} = 30$ $h^{-1}$ Mpc is used \cite{KP4s4-Paillas}, rather than ``RecIso'' \cite{Padmanabhan12} for all tracers, allowing a much simpler model for the post-reconstruction correlation function with only a single Gaussian damping terms with line-of-sight and transverse damping scales. Rather than tying together the damping terms with $\Sigma_{\textrm{sm}}$ as in \cite{BOSSBAOAmp}, we now 
allow $\Sigma_{\perp}$ and $\Sigma_{\parallel}$ to vary separately
with broad flat priors. 

\section{Mock tests}

\subsection{Testing the EFT model fit}
\label{sec:full_shape_mock_tests}

We refer the reader to our companion paper \cite{Crespi25}, for a battery of extensive tests of the new method, investigating the recovered parameters for many mock power spectra. We highlight a few of the tests here by showing, in Fig.~\ref{fig:abacus_mock_test}, that we can accurately recover $\gamma_b$ (and thus the density-derived Hubble parameter $h^{\textrm{dens}}$) when fitting to the mean of the 25 Abacus mocks. As detailed in Table~\ref{tab:full_shape_summary_main}, we find biases of $<0.3\sigma$ in both $h^{\textrm{dens}}$ and $\gamma_b$, and good agreement between the marginalized mean and the maximum of the posterior. We also find that the spread among the 25 Abacus mocks is consistent with the reported errors.

\begin{table*}[]
    \centering
    \begin{tabular}{l|cc|cc|cc}
    & \multicolumn{2}{c|}{Truth} & \multicolumn{2}{c|}{$h$} &  \multicolumn{2}{c}{$\gamma_b$} \\
    Mock & $h$ & $f_b$ & Mean $\pm$ 1$\sigma$ (MAP) & $n\sigma$ &  Mean $\pm$ 1$\sigma$ (MAP) & $n\sigma$ \\
    \hline
     Abacus & 0.6736 & 0.1571 & $0.663 \pm 0.033$ (0.662) & -0.30 & $0.160 \pm 0.011$ (0.160) & 0.26 \\
    $\Lambda$CDM Noiseless & 0.6736 & 0.1571 & $0.681 \pm 0.037$ (0.677) & 0.20 & $0.155 \pm 0.0144$ (0.155)  & -0.14 \\
    EDE Noiseless & 0.7219 & 0.1471 & $0.721 \pm 0.040$ (0.716)  & -0.03 & $0.1479 \pm 0.015$ (0.150) & 0.05 \\
    \end{tabular} 
    \caption{Summary of full-shape fits to the mean of the 25 Abacus mocks and noiseless theory vectors in both the default $\Lambda$CDM and early dark energy cosmologies.
    \label{tab:full_shape_summary_main}}
\end{table*}

\begin{figure}
    \includegraphics[width=0.5\textwidth]{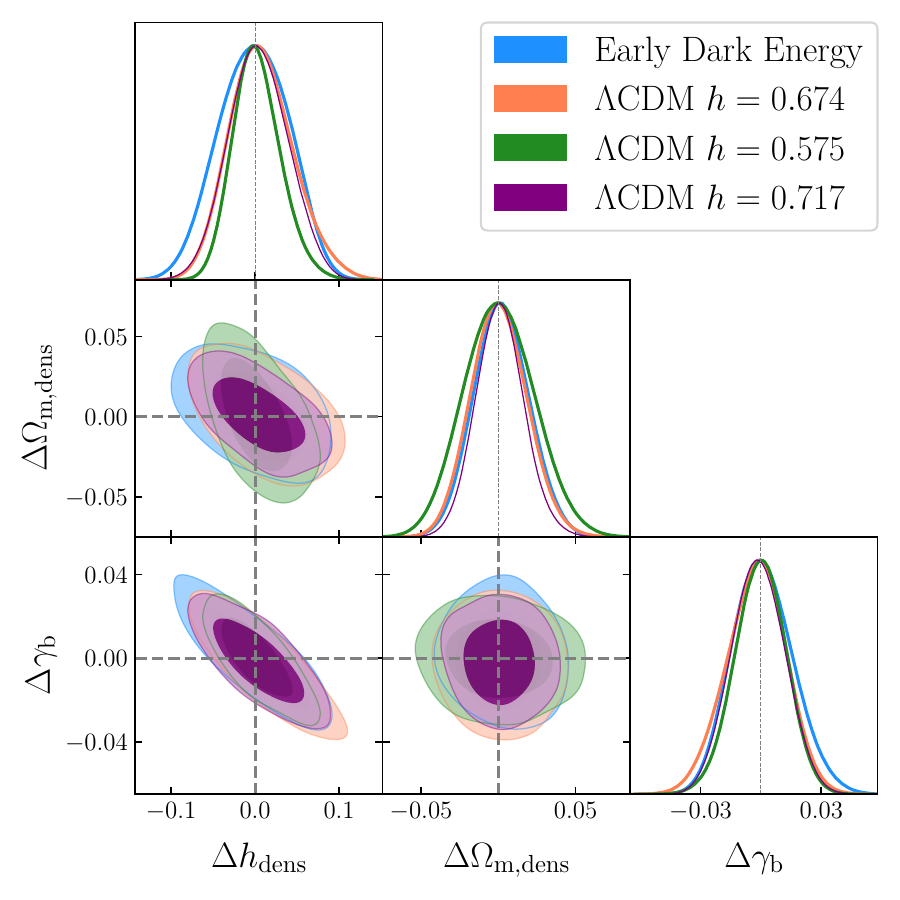}
    \caption{Our method accurately recovers the Hubble parameter
    across a wide range in true $h$, and in both a $\Lambda$CDM and an Early Dark Energy cosmology (with $h = 0.722$).
    \label{fig:different_cosmologies_contours}}
\end{figure}

\begin{figure}
    \includegraphics[width=0.5\textwidth]{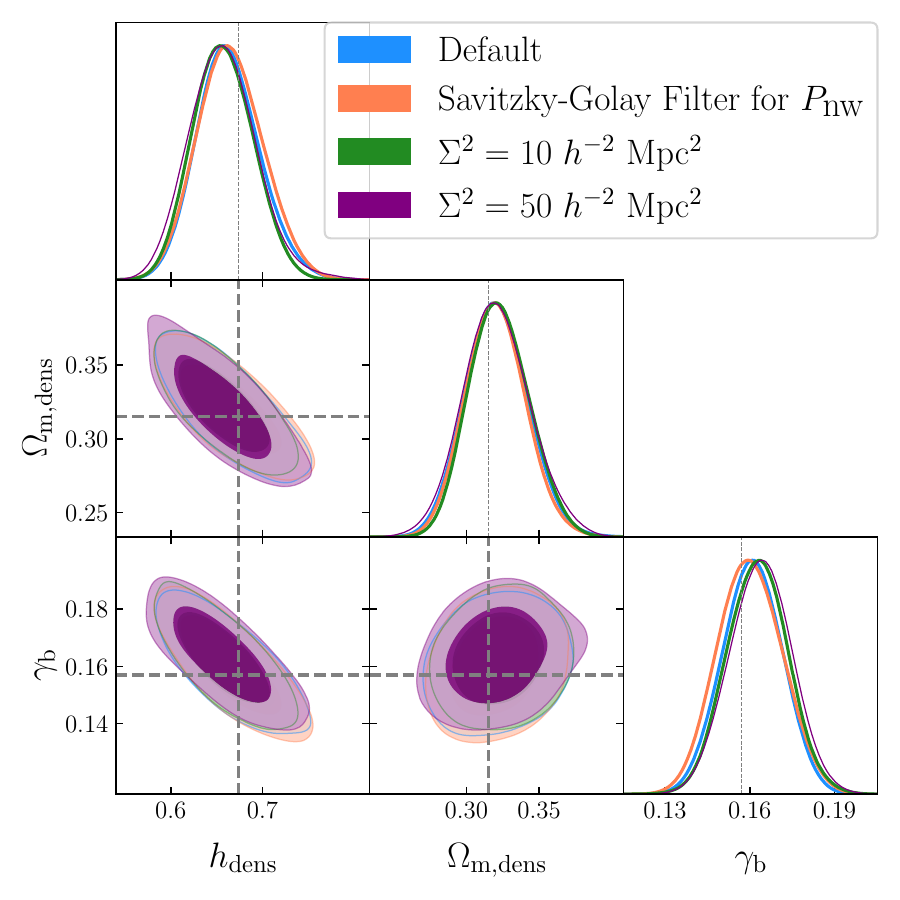}
    \caption{The recovery of $\gamma_b$ and $h^{\textrm{dens}}$ on the pre-reconstruction power spectrum is not sensitive to the details of IR resummation. Here we compare the default result (on the mean of the 25 Abacus mocks) to runs where we change the filter used to construct the no-wiggle power spectrum (orange) or those where we change the damping of the wiggle power spectrum (green and purple).
    \label{fig:ir_resummation_test}}
\end{figure}



We have also verified our recovery of the baryon fraction on noiseless theory vectors spanning a wide range in input $\Lambda$CDM and Early Dark Energy cosmologies. As shown in Fig.~\ref{fig:different_cosmologies_contours}, our method recovers the correct baryon fraction in a $\Lambda$CDM cosmology generated in the DESI fiducial cosmology, an Early Dark Energy cosmology with $h = 0.722$, and $\Lambda$CDM cosmologies with both high and low values of $h$. 
We show more extensive tests in \cite{Crespi25}, where we show that $h$ and $\gamma_b$ are biased by $<0.1\sigma$ in both $\Lambda$CDM and EDE cosmologies across a broad range in true $h$, $0.55 < h < 0.8$.

In Fig.~\ref{fig:ir_resummation_test}, we test the sensitivity of our results on the mean of the Abacus mocks to changing the IR resummation parameters. We test both changing the method for the IR resummation split (from the default method to separate the wiggle and no-wiggle power spectrum from \cite{Wallisch18}, to using a Savitzky-Golay filter to define the no-wiggle power spectrum, following tests in \cite{KP4s2-Chen}), and fixing the damping of the wiggle power spectrum \cite{Chen20} to either $\Sigma^2 = 10$ $h^{-2}$ Mpc$^2$ or $\Sigma^2 = 50$ $h^{-2}$ Mpc$^2$---roughly bracketing the default value of $\Sigma^2 \sim 30$ for all tracers in the default DESI cosmology. We find that these changes minimally affect our results, changing $H_0$ by $<0.25\sigma$, and conclude that our method is robust to the details of the IR resummation method used.

\begin{table*}[]
    \centering
    \begin{tabular}{l|ccc}
    Parameter & Truth & Fit to mean & Med.\ of 25 fits \\
    \hline
    BGS, $0.1 < z < 0.4$ \\
    \hline
    $\alpha_{\textrm{iso}}$ & 1.0067 & $1.0057 \pm 0.0226$ & $1.0021 \pm 0.0265$ \\
    $\gamma_b$ & 0.1571 & $0.1362 \pm 0.0449$ & $0.1379 \pm 0.0505$ \\
    $\chi^2$ & 29 & 1.0 & 28.3 \\    
    \hline
    LRG, $0.4 < z < 0.6$ \\
    \hline
    $\alpha_{\textrm{iso}}$ & 1.006 & $1.0068 \pm 0.0139$ & $1.0121 \pm 0.0154$ \\
    $\alpha_{\textrm{AP}}$ & 0.998 & $0.9990 \pm 0.0534$ & $1.0064 \pm 0.0547$ \\
    $\gamma_b$ & 0.1571 & $0.1633 \pm 0.0358$ & $0.1682 \pm 0.0360$ \\
    $\chi^2$ & 59 & 3.7 & 53.4 \\
    \hline
    LRG, $0.6 < z < 0.8$ \\
    \hline
    $\alpha_{\textrm{iso}}$ & 1.0050 & $1.0056 \pm 0.0112$ & $1.0045 \pm 0.011$ \\
    $\alpha_{\textrm{AP}}$ & 0.9978 & $1.0082 \pm 0.0432$ & $1.0084 \pm 0.0392$ \\
    $\gamma_b$ & 0.1571 & $0.1574 \pm 0.0354$ & $0.1554 \pm 0.0331$ \\
    $\chi^2$ & 59 & 3.3 & 64.3 \\
    \hline
    LRG+ELG, $0.8 < z < 1.1$  \\
    \hline
    $\alpha_{\textrm{iso}}$ & 1.0044 & $1.0068 \pm 0.0088$ & $1.0082 \pm 0.0085$ \\
    $\alpha_{\textrm{AP}}$ & 0.9974 & $1.0028\pm 0.0293$ & $0.9934 \pm 0.0277$ \\
    $\gamma_b$ & 0.1571 & $0.1638 \pm 0.0262$ & $0.1559 \pm 0.0262$ \\
    $\chi^2$ & 59 & 4.3 & 60.2 \\
    \hline
    ELG, $1.1 < z < 1.6$ \\
    \hline
    $\alpha_{\textrm{iso}}$ & 1.0036 & $1.0049 \pm 0.0155$ & $1.0068 \pm 0.0160$ \\
    $\alpha_{\textrm{AP}}$ & 0.9971 & $0.9934 \pm 0.0572$ & $1.0008 \pm 0.0555$ \\
    $\gamma_b$ & 0.1571 & $0.1608 \pm 0.0389$ & $0.1610 \pm 0.0372$ \\
    $\chi^2$ & 59 & 5.8 & 57.9 \\
    \hline
    QSO, $0.8 < z < 2.1$ \\
    \hline
    $\alpha_{\textrm{iso}}$ & 1.0033 & $1.0065 \pm 0.0178$ & $0.9961 \pm 0.0179$ \\
    $\gamma_b$ & 0.1571 & $0.1585 \pm 0.0359$ & $0.1529 \pm 0.0377$ \\
    $\chi^2$ & 29 & 1.8 & 34.4
    \end{tabular}
    \caption{Recovery of BAO scaling parameters and $\gamma_b$ on the post-reconstruction correlation function of 25 Abacus altmtl cutsky mocks, analyzed in the BOSS fiducial cosmology, and using the DESI data covariance. The first column gives the true value of each parameter, or the number of degrees of freedom for the dataset. The second column gives the result of a fit to the mean of the 25 mocks, using the DESI data covariance matrix. The final column gives the median and median of the errorbar of fits to each of the 25 different mocks.
    The fits to the mean have smaller $\chi^2$ because we use the DESI DR1 covariance matrix everywhere.
    Combining the fits to the mean from each tracer gives $\gamma_b = 0.159 \pm 0.014$.
    \label{tab:abacus_post_recon_tests}}
\end{table*}


\begin{figure*}
\includegraphics[width=\textwidth]{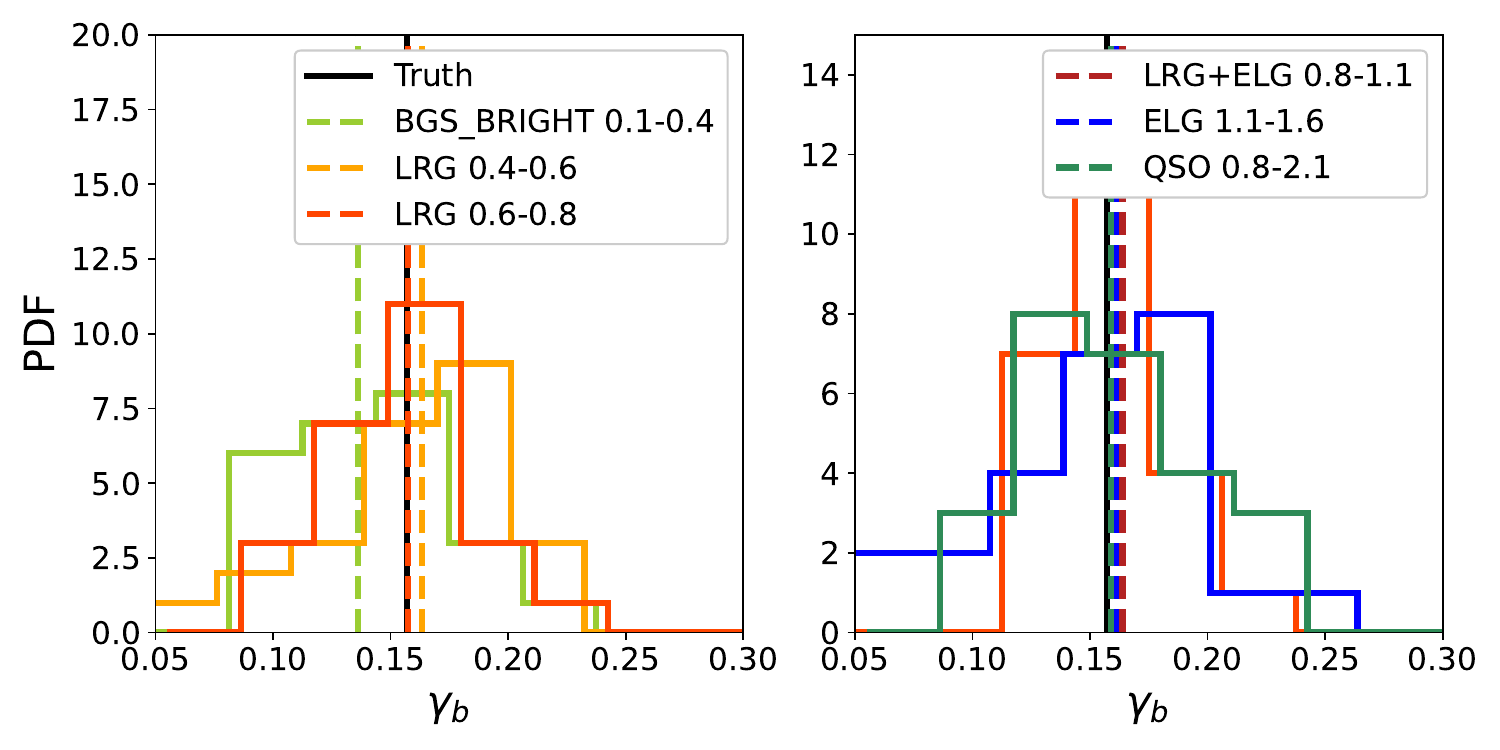}
\caption{Distribution of $\gamma_b$ fit to the post-reconstruction correlation function of all DESI tracers in the 25 Abacus mocks; the tracers are split between the two panels for clarity. The solid black line shows the true value of the baryon fraction and the dashed lines show the fits to the mean of the 25 Abacus mocks.
\label{fig:post_recon_gamma_B}} 
\end{figure*}

\begin{figure*}
\includegraphics[width=\textwidth]{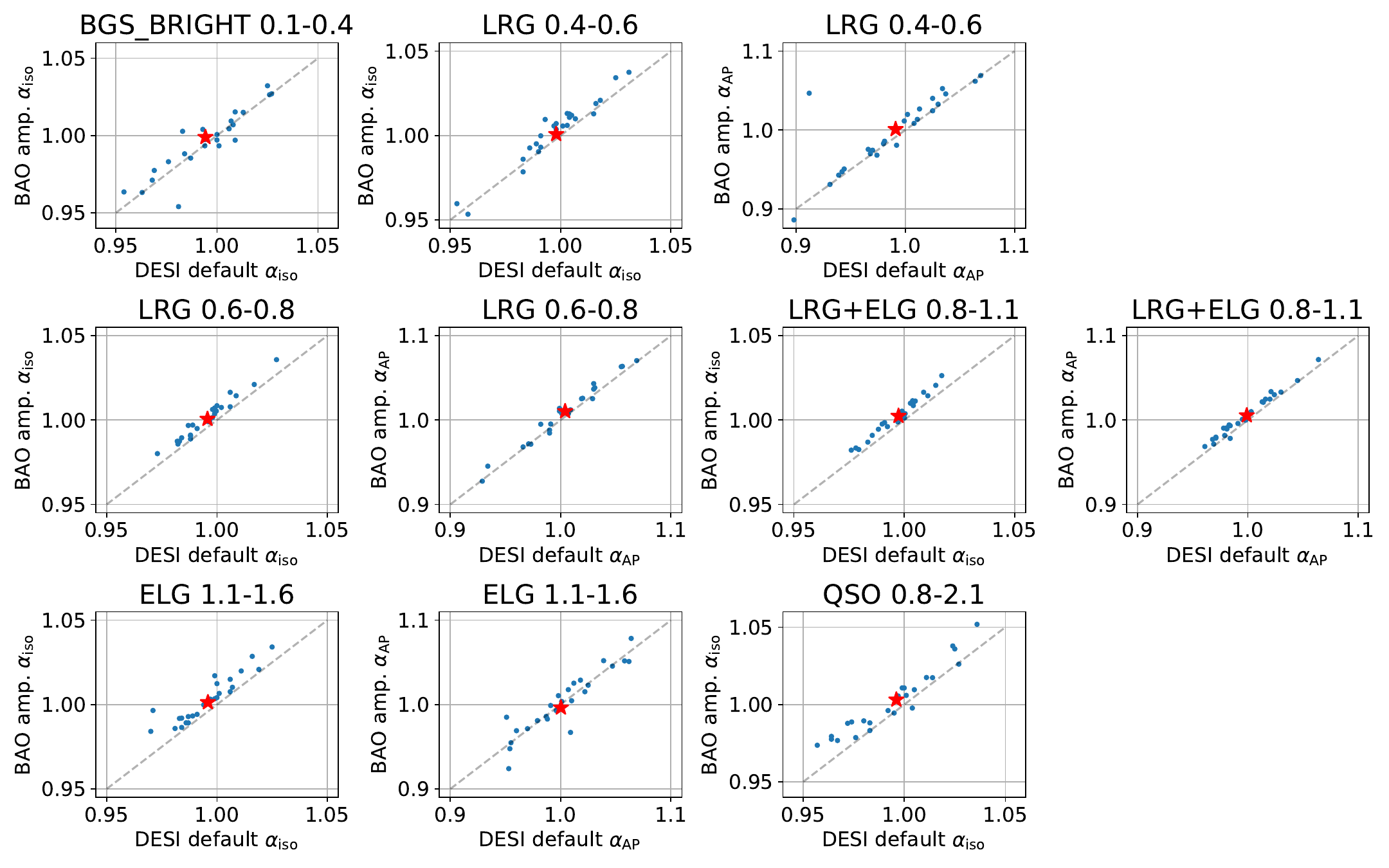}
\caption{Comparison between Alcock-Paczynski parameters, $\alpha_{\textrm{iso}}$ and $\alpha_{\textrm{AP}}$, from the default DESI fits to the post-reconstruction correlation function, and
fit using our pipeline with $\gamma_b$ as a free parameter. Blue dots show the fit to each of the 25 Abacus mocks, while the red star shows the fit to the mean. 
The BAO amplitude fits have been rescaled to match the DESI fiducial cosmology.
The dashed line indicates perfect agreement between the two pipelines. 
\label{fig:post_recon_alpha_comparison}} 
\end{figure*}

\begin{figure}
\includegraphics[width=0.5\textwidth]{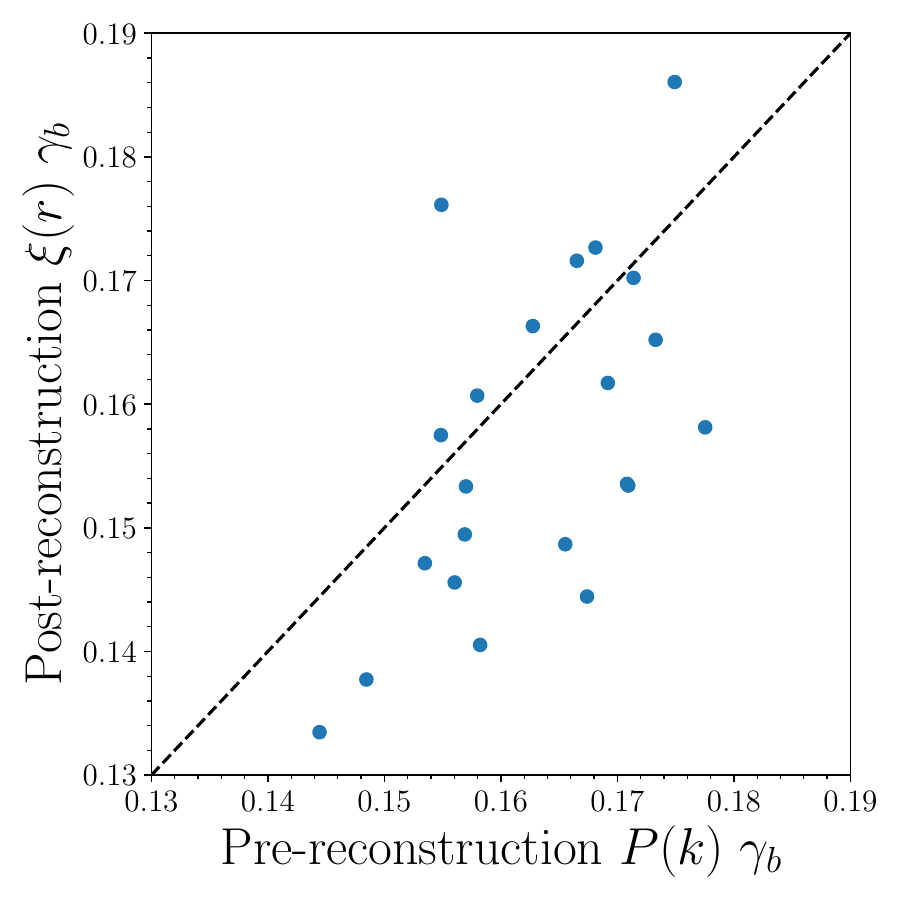}
\caption{Comparison between $\gamma_b$ measurement from the pre-reconstruction power spectrum and the post-reconstruction correlation function for the 25 Abacus mocks.
\label{fig:pre_vs_post_abacus}} 
\end{figure}

\subsection{Testing the post-reconstruction BAO fits}
\label{sec:BAO_mock_tests}


Given the changes in our pipeline fitting $\gamma_b$ to the post-reconstruction correlation function, we have also undertaken a series of tests of the revised method.
We find that our pipeline 
accurately recovers the true value of $\gamma_b$ (Fig.~\ref{fig:post_recon_gamma_B}) and matches the DESI DR1 fits to the Abacus mocks well (Fig.~\ref{fig:post_recon_alpha_comparison}). 
We compare the BAO parameters to the DESI DR1 BAO fits in more detail in Appendix~\ref{sec:post_recon_compare}, showing that the constraining power of our pipeline on $\alpha_{\textrm{iso}}$ is similar, but the $\alpha_{\textrm{AP}}$ constraints are slightly worse due to the larger number of free parameters in our model.
We summarize the fits to each tracer in Table~\ref{tab:abacus_post_recon_tests}, and show that the model fits the data well with $\chi^2$ matching the expected number of degrees of freedom.
The small residual differences between our results and the DESI DR1 pipeline can be attributed to slightly different nuisance parameter priors. In particular, our pipeline has slightly larger
errors on $\alpha_{\textrm{AP}}$ due to the broad priors on the damping parameters $\Sigma_{\parallel}$, $\Sigma_{\perp}$, and $\Sigma_{\textrm{fog}}$.


Combining the fits to the mean of the 25 Abacus mocks, we find $\gamma = 0.159 \pm 0.0154$, a 0.11$\sigma$ bias away from the truth in units of the DR1 errorbar. Since we have 25 Abacus mocks, the error on the mean is 5 times smaller, meaning that the bias is statistically consistent with zero.

We fit cosmological parameters to the compressed results from Table~\ref{tab:abacus_post_recon_tests}, using the DESI DR1 measurements and covariance for $\alpha_{\textrm{iso}}$ and $\alpha_{\textrm{AP}}$.
We measure the cross-covariance between $\gamma_b$, $\alpha_{\textrm{iso}}$ and $\alpha_{\textrm{AP}}$ from the parameter covariance in our fits to the post-reconstruction correlation function.
We find $h^{\textrm{dens}} = 0.679^{+0.044}_{-0.037}$ and $\gamma_b = 0.158 \pm 0.014$, in good agreement with the full-shape results although with less constraining power, as expected since we are not using any shape information.
This is consistent with our previous results on BOSS, which showed that reconstruction changes the BAO damping parameter but does not tighten constraints on $\gamma_b$ \cite{BOSSBAOAmp}.

In Fig.~\ref{fig:pre_vs_post_abacus}, we compare the pre-reconstruction and post-reconstruction fits for $\gamma_b$ on the 25 Abacus mocks. As expected, they are highly correlated, with a correlation coefficient of 0.6. We tested including the post-reconstruction $\gamma_b$ in our data vector along with the post-reconstruction BAO parameters, using the 1000 EZmocks to measure the cross-covariance with the pre-reconstruction power spectrum. We find that this leads to only a 6\% improvement in the constraining power on $H_0$, consistent with the strong correlation between pre and post-reconstruction $\gamma_b$. Due to the minimal gain in constraining power and to avoid additional complexity regarding the impact of reconstruction on $\gamma_b$, we do not choose to combine the pre- and post-reconstruction $\gamma_b$ by default.



\section{Results and discussion}
\label{sec:data}

\begin{figure*}
\includegraphics[width=\textwidth]{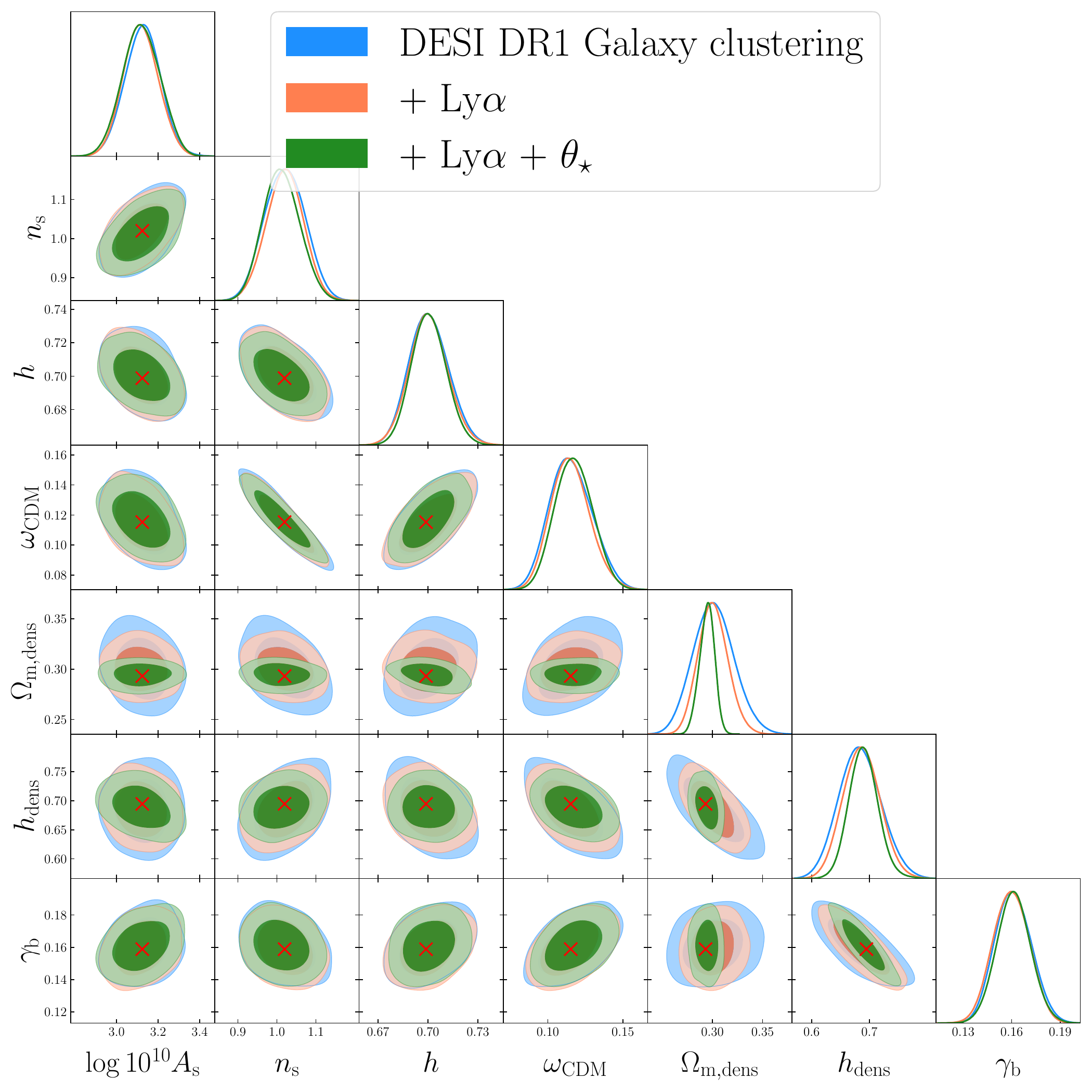}
\caption{Constraints on density-derived $H_0$ using DESI DR1 galaxy clustering data (blue), also including DESI DR1 Ly$\alpha$ (orange), and finally when also including Planck $\theta_\star$ to improve the constraints on $\Omega_{\mathrm{m}}^{\textrm{dens}}$ (green). The MAP point (when including Ly$\alpha$ and $\theta_\star$) is shown with red crosses.
\label{fig:desi_data_results}} 
\end{figure*}

\begin{table*}[]
    \centering
    \begin{tabular}{l|ccc}
    Data combination & $H_0^{\textrm{dens}}$ (km s$^{-1}$ Mpc$^{-1}$) & $\Omega_{\mathrm{m,dens}}$ & $\gamma_b$ \\
    \hline
    DESI DR1 Galaxy clustering (Full Shape $\gamma_b$) & $68.4 \pm 3.6$ & $0.301 \pm 0.020$ & $0.161 \pm 0.010$ \\
    DESI DR1 Galaxy clustering (Post-recon $\gamma_b$) & $70.7 \pm 4.5$ & $0.294 \pm 0.022$ & $0.153 \pm 0.014$ \\
    DESI DR1 Galaxy clustering  + Ly$\alpha$ & $68.6 \pm 3.1$ & $0.301 \pm 0.014$ & $0.161 \pm 0.010$ \\
    DESI DR1 Galaxy clustering + Ly$\alpha$ + $\theta_{\star}$ & $69.0 \pm 2.5$ & $0.294 \pm 0.007$ & $0.161 \pm 0.010$ \\
    \end{tabular} 
    \caption{Constraints on density-derived cosmological parameters from DESI DR1. Top row shows our fiducial results, measuring $\gamma_b$ from EFT modelling of the power spectrum, using DESI DR1 galaxy clustering data only. The middle row shows the results when we only use BAO amplitude information, measuring $\gamma_b$ in the post-reconstruction correlation function and marginalizing over all shape information. The bottom row shows the fiducial method combined with DESI DR1 Ly$\alpha$ and Planck $\theta_\star$.
    \label{tab:desi_data_results}}
\end{table*}

\begin{table*}[]
    \centering
    \begin{tabular}{l|ccc}
    Tracer & $\gamma_b$ & $\alpha_{\textrm{iso}}$ & $\alpha_{\textrm{AP}}$ \\
    \hline
    BGS\_BRIGHT 0.1-0.4 & $0.106\pm0.050$ & $0.981 \pm0.026$ \\
    LRG 0.4-0.6 & $0.146^{+0.034}_{-0.037}$ & $0.986 \pm 0.013$ & $0.923 \pm 0.040$ \\ 
    LRG 0.6-0.8 & $0.219^{+0.030}_{-0.036}$ & $0.967 \pm 0.010$ & $1.036 \pm 0.039$ \\
    LRG+ELG 0.8-1.1 & $0.157\pm0.030$ & $1.005 \pm 0.008$ & $1.027 \pm 0.028$  \\
    ELG 1.1-1.6 & $0.156^{+0.041}_{-0.036}$ & $0.998 \pm 0.014$ & $0.971 \pm 0.043$ \\
    QSO 0.8-2.1 & $0.122\pm0.037$ & $1.017 \pm 0.0250$
    \end{tabular} 
    \caption{Results on data from post-recon fits to each tracer. $\alpha_{\textrm{iso}}$ and $\alpha_{\textrm{AP}}$ are quoted relative to the DESI fiducial cosmology.
    \label{tab:post_recon_by_tracer}}
\end{table*}

After passing the validation tests described in \cite{Crespi25} and in Sections~\ref{sec:full_shape_mock_tests} and~\ref{sec:BAO_mock_tests}, we unblinded and performed the same fits on the data.
Our fiducial full-shape based results are shown in Fig.~\ref{fig:desi_data_results}
and summarized in Table~\ref{tab:desi_data_results}.

Using the DESI DR1 galaxy clustering data for the baryon fraction and $\Omega_{\mathrm{m}}$, and the latest BBN constraint on $\Omega_{\mathrm{b}} h^2 = 0.02218 \pm 0.00055$ from \cite{Schoneberg24}, we find $H_0 = 68.4 \pm 3.6$ km s$^{-1}$ Mpc$^{-1}$. 

We find consistency between our fiducial method, inferring the baryon fraction from the full-shape of the power spectrum, and the alternative where we measure the baryon fraction from the post-reconstruction correlation function. Specifically, the difference in $H_0$ between the two methods (2.3 km s$^{-1}$ Mpc$^{-1}$) is consistent with the scatter expected due to the excess variance in the post-reconstruction method compared to the full-shape method (2.7 km s$^{-1}$ Mpc$^{-1}$).\footnote{Determined as $\sqrt{4.5^2 - 3.6^2}$, the quadrature difference of the $H_0$ errors from the post-reconstruction $\gamma_b$ and the full-shape $\gamma_b$} The full-shape method gives tighter errors on $\gamma_b$ (and thus $H_0$), as expected because it incorporates baryon-fraction information from the shape of the power spectrum that is marginalized over by the broadband parameters in the post-reconstruction fits. $\Omega_{\mathrm{m,dens}}$ is slightly different between the two methods because in the full-shape fits, we use the 
power spectrum and post-reconstruction AP parameters from the LRG3 bin at $0.8 < z < 1.1$, whereas in the post-reconstruction fits, we use the combined LRG+ELG correlation function at $0.8 < z < 1.1$.

When including DESI DR1 Ly$\alpha$ BAO, we find for our baseline DESI-only constraints, $H_0 = 68.6 \pm 3.1$ km s$^{-1}$ Mpc$^{-1}$.
We can further improve the constraints by also including Planck  PR3 $\theta_\star$, effectively an additional angular BAO measurement at $z \sim 1090$.
While there is some disagreement between Planck PR3 and DESI DR1 in the $\Omega_{\mathrm{m}}$-$H_0 r_d$ plane, the tension is $<2 \sigma$, and it is worse in the $H_0 r_d$ direction ($1.8\sigma$) than $\Omega_{\mathrm{m}}$ (1.2$\sigma$). It is therefore justified to combine these datasets.

When combining with $\theta_\star$, we find $H_0 = 69.0 \pm 2.5$ km s$^{-1}$ Mpc$^{-1}$.
This is in good agreement (0.7$\sigma$ tension) with the CMB constraints on $H_0$ in $\Lambda$CDM ($67.24 \pm 0.35$ km s$^{-1}$ Mpc$^{-1}$ from \cite{SPT25}). Our measurement is also consistent with the local $H_0$ measurement from SH0ES ($73.04 \pm 1.04$ km s$^{-1}$ Mpc$^{-1}$) \cite{Riess-Hubble}, with a slightly larger difference of 1.5$\sigma$.

In our previous work, we used Type Ia supernovae as uncalibrated standard candles to further improve the geometric constraints on $\Omega_{\mathrm{m}}$. However, the $\Omega_{\mathrm{m}}$ constraints from the supernovae are in modest tension with those from DESI and $\theta_\star$ (1.9$\sigma$ for Pantheon+, 2.2$\sigma$ for Union3, and 3.5$\sigma$ for DESY5SN). 
This tension arises from differences in the $z<0.5$ distance scale preferred by  supernovae vs.\ BAO and CMB, and the physical origin is unclear: it could arise from systematics in one or more samples, new physics, or possibly just be a large statistical fluctuation.
One possible explanation is the 2-3$\sigma$ preference for $w_0 w_a$CDM over $\Lambda$CDM when combining DESI DR1, CMB, and supernovae. Given that we infer $\Omega_{\mathrm{m}}$ from geometric probes assuming flat $\Lambda$CDM, we therefore opt not to combine with supernovae in our baseline constraints, which may not be consistent with this model.

If we did use the supernovae data, and inferred $\Omega_{\mathrm{m}}$ in an evolving dark energy cosmology, motivated by the tensions in $\Omega_{\mathrm{m}}$ between supernovae and other probes, we would find $\Omega_{\mathrm{m}} = 0.323 \pm 0.0095$ using Union3 supernovae, from Table 3 in \cite{DESI2024.III.KP4}. The larger $\Omega_\mathrm{m}$ leads to a smaller $H_0 \propto \Omega_{\mathrm{m}}^{-0.5}$ (from Eq.~\ref{eq:basic_eqn}), and applying this scaling leads to a best-fit $H_0$ = 66.0 km s$^{-1}$ Mpc$^{-1}$, still in $\sim$2$\sigma$ agreement with SH0ES.
When combining with other supernovae samples, $\Omega_{\mathrm{m}}$ is closer to the best-fit DESI plus $\theta_\star$ value, and this leads to smaller shifts in $H_0$, to 66.7 km s$^{-1}$ Mpc$^{-1}$ with DESY5 or 67.5 km s$^{-1}$ Mpc$^{-1}$ with Pantheon+.
While moving to a $w_0$-$w_a$ cosmology reduces constraining power on $\Omega_{\mathrm{m}}$, the extra information from the supernovae compensates for this reduction and thus the $\Omega_{\mathrm{m}}$ error is similar to our baseline constraints including $\theta_\star$.

A nonzero spatial curvature could also mitigate the discrepancy between DESI and the CMB \cite{Chen25}. When $\Omega_k$ is free, the $\Omega_{\mathrm{m}}$ constraints from DESI + BBN + $\theta_\star$ become considerably weaker ($0.296 \pm 0.014$) but tighten when adding the full CMB likelihood ($0.305 \pm 0.0051$) \cite{DESI2024.III.KP4}. The small shift in $\Omega_{\mathrm{m}}$ in a free curvature model would shift $H_0$ down to $\sim$ 68 km s$^{-1}$ Mpc$^{-1}$.

Freeing the neutrino mass does not change the inference of $\Omega_{\mathrm{m}}$ much. The DESI + CMB $\Omega_{\mathrm{m}}$ measurement changes from $0.3069 \pm 0.0050$ in $\Lambda$CDM to $0.3037 \pm 0.0053$ in $\nu\Lambda$CDM. Likewise, in $w_0 w_a$CDM (using Union3 supernovae, but results are similar for DES Y5 and Pantheon+), $\Omega_{\mathrm{m}}$ changes from $0.323 \pm 0.0095$ to $0.324 \pm 0.0098$ when freeing the neutrino mass \cite{DESI2024.III.KP4}.

\section{Conclusions and outlook}

We have applied a new method to infer the Hubble constant from the total energy density of the Universe, independent of standard rulers to DESI DR1 data. This method relies on comparing the physical matter density $\epsilon_m$ (from the CMB temperature, BBN, and the baryon fraction from galaxy clustering) with the fractional matter density $\Omega_{\mathrm{m}}$ (from relative sizes of geometric probes).
We show that this method is robust when applied both to the DESI DR1 Abacus mocks, and to noiseless theory vectors drawn from $\Lambda$CDM and Early Dark Energy cosmologies across a wide range in $H_0$. This explicitly shows that this method can be used to test models like EDE that resolve the Hubble tension by modifying the sound horizon.

We validate measurements of the baryon fraction both using the full-shape of the power spectrum fit with the EFT of Large Scale Structure (our fiducial method) and fitting the amplitude of the BAO feature in the post-reconstruction correlation function. For both methods, we have developed a new, more consistent method to consistently extract baryon fraction information from the dependence of the power spectrum on the gravitational component of the growth of structure (as described in our companion paper \cite{Crespi25}), independent of the baryon fraction information in the background or thermal history. Both methods return consistent $H_0$ constraints both on mocks and on data, with our fiducial method returning tighter constraints as it includes full-shape information that is marginalized in the post-reconstruction BAO fitting.

When inferring $\Omega_{\mathrm{m}}$ from the relative BAO positions in DESI and the CMB (i.e.\ $\theta_\star$), we find $H_0 = 69.0 \pm 2.5$ km s$^{-1}$ Mpc$^{-1}$. This is consistent with both the Planck and SH0ES $H_0$ measurements, though closer to the Planck value. 

The constraining power of our measurement is similar to other $r_d$-marginalized measurements from DESI DR1. Using the DESI DR1 power spectrum and CMB lensing, \cite{Zaborowski24} find $H_0 = 70.1^{+2.7}_{-3.3}$ km s$^{-1}$ Mpc$^{-1}$, and $H_0 = 66.7^{+1.7}_{-1.9}$ km s$^{-1}$ Mpc$^{-1}$ when also using $\Omega_{\mathrm{m}}$ information from DES-Y5 supernovae. Using the same methodology but replacing supernovae with DESI DR1 BAO and Planck $\theta_\star$, \cite{Zaborowski25} finds $H_0 = 69.2^{+1.4}_{-1.3}$ km s$^{-1}$ Mpc$^{-1}$.
Similarly, using DESI DR2, CMB lensing, galaxy lensing, and supernovae, \cite{GarciaEscudero25} finds $H_0 = 70 \pm 1.7$ km s$^{-1}$ Mpc$^{-1}$. However, our method is more explicitly independent of pre-recombination $H_0$ modifications than these other works. The sound-horizon-marginalized method of \cite{Zaborowski24,Zaborowski25} infers $H_0$ by comparing $\Omega_{\mathrm{m}}$ to $\Omega_{\mathrm{m}} h^2$ inferred from the shape of the power spectrum (i.e.\ setting the power spectrum turnover scale, though they do not explicitly measure the turnover as they use $k_{\textrm{min}} = 0.02$ $h$ Mpc$^{-1}$). However, new physics that alters the sound horizon can also change the broadband shape of the power spectrum, affecting inference of $\Omega_{\mathrm{m}} h^2$ and thus $H_0$ \cite{Smith22}. 
In contrast, we show explicitly that our method recovers the correct value of $H_0$ in early dark energy cosmologies.

Our method is currently limited by the accuracy of the baryon fraction measurement; however, this is expected to improve greatly with current and future DESI data. The final DESI data will increase the sky coverage by 3-4$\times$, as well as increasing completeness and number density (particularly critical for the dense ELG sample). 
The constraining power of energy-density based $H_0$ from DESI DR1 is already $\sim2\times$ better than the constraining power from BOSS, and the considerably larger effective volume of the final DESI survey should allow for another $\sim2\times$ gain in precision. One potential limitation is the baryon-CDM relative velocity effect; we are developing methodology to jointly fit the relative velocity biases and the BAO amplitude, and we could further tighten the constraints on the relative velocity biases using the bispectrum \citep{Yoo11,Slepian18} or expectations on the relative velocity biases from simulations \citep{Barreira20,Khoraminezhad20}.
These significant improvements will enable this measurement to differentiate between the local and cosmological $H_0$ measurements with the final data from DESI.

\section*{Data Availability}
The data used in this work are public as part of DESI Data Release 1 (details at \url{https://data.desi.lbl.gov/doc/releases/}). The data points corresponding to the figures
are available on Zenodo at
\url{https://zenodo.org/uploads/17686137}. Code is available on GitHub at \url{https://github.com/drew2799/fbCAMB}.

\acknowledgments

AK was supported as a CITA National Fellow by the Natural Sciences and Engineering Research Council of Canada (NSERC), funding reference \#DIS-2022-568580.
WP acknowledges support from the Natural Sciences and Engineering Research Council of Canada (NSERC), [funding reference number RGPIN-2025-03931] and from the Canadian Space Agency.
Research at Perimeter Institute is supported in part by the Government of Canada through the Department of Innovation, Science and Economic Development Canada and by the Province of Ontario through the Ministry of Colleges and Universities.
This research was enabled in part by support provided by Compute Ontario (computeontario.ca) and the Digital Research Alliance of Canada (alliancecan.ca).

This material is based upon work supported by the U.S. Department of Energy (DOE), Office of Science, Office of High-Energy Physics, under Contract No. DE–AC02–05CH11231, and by the National Energy Research Scientific Computing Center, a DOE Office of Science User Facility under the same contract. Additional support for DESI was provided by the U.S. National Science Foundation (NSF), Division of Astronomical Sciences under Contract No. AST-0950945 to the NSF’s National Optical-Infrared Astronomy Research Laboratory; the Science and Technology Facilities Council of the United Kingdom; the Gordon and Betty Moore Foundation; the Heising-Simons Foundation; the French Alternative Energies and Atomic Energy Commission (CEA); the National Council of Humanities, Science and Technology of Mexico (CONAHCYT); the Ministry of Science, Innovation and Universities of Spain (MICIU/AEI/10.13039/501100011033), and by the DESI Member Institutions: \url{https://www.desi.lbl.gov/collaborating-institutions}. Any opinions, findings, and conclusions or recommendations expressed in this material are those of the author(s) and do not necessarily reflect the views of the U. S. National Science Foundation, the U. S. Department of Energy, or any of the listed funding agencies.

The authors are honored to be permitted to conduct scientific research on I'oligam Du'ag (Kitt Peak), a mountain with particular significance to the Tohono O’odham Nation.

Software: \texttt{getdist} \cite{GetDist},
\texttt{desilike}\footnote{\url{https://desilike.readthedocs.io/en/latest/}}, \texttt{velocileptors} \cite{Chen20,Chen21}, \texttt{Julia}, \texttt{Lux.jl}~\cite{pal2023lux}, \texttt{Turing.jl}~\cite{turing1, turing2}, \texttt{Effort.jl}~\cite{Effort}, \texttt{CLASS-PT} \cite{Chudaykin20}, \texttt{baofit}\footnote{\url{https://github.com/ashleyjross/BAOfit}}.

\bibliographystyle{JHEP}
\bibliography{main,DESI2024}

\appendix


\section{Impact of relative velocity}
\label{sec:relative_velocity}

Besides perturbations in the total combined density of baryons and dark matter, galaxy clustering
can also respond to relative perturbations in the density and velocity of baryons and dark matter \cite{Schmidt16,Chen19a,KP4s2-Chen}. These relative perturbations grow much slower than the standard adiabatic
growing mode, but they have a larger imprint of the BAO feature and hence are a potential source of systematic error for this work.
At linear order, the relative density perturbations decouple from the matter perturbations and grow as \cite{Schmidt16}
\begin{equation}
    \delta_{\textrm{bc}}(a) = R_+ + R_- D_{\textrm{bc}}(a)
\end{equation}
where $R_+$ and $R_-$ are constant in redshift and 
\begin{equation}
    D_{\textrm{bc}} = \int_{\ln{a}}^\infty \frac{d \ln a'}{a'^2 H(a')/H_0}
\end{equation}
The solution for relative density perturbations thus consists of two modes, one that is a constant compensated perturbation where $\delta_c$ and $\delta_b$ oppose each other, and another decaying mode
that arises from the initial relative velocity, $v_{bc}$, which decays as $1/a$:
\begin{equation}
    R_- = \frac{\theta_{\textrm{bc}}(z=0)}{H_0}
\end{equation}
where $\theta_{\textrm{bc}}$ is the divergence of the relative  velocity.The linear evolution of $\delta_{\textrm{bc}}$ and $\theta_{\textrm{bc}}$ can therefore be computed from the transfer functions $T_{\textrm{b}}(k)$, $T_{\textrm{c}}(k)$. $R_-(k)$ and $R_+(k)$ are given by
\begin{equation}
    R_-(k) = \frac{k}{H_0} \frac{T_{\textrm{vbc}}(k)}{T_m(k)} \delta_m(k, z=0)
\end{equation}
\begin{equation}
R_+(k) = \frac{T_b(k) - T_c(k)}{T_m(k)} \delta_m(k,0) - R_-(k) D_{\textrm{bc}}(z=0)
\end{equation}
where the second term is $\sim$1\% of the first term.

\begin{figure*}
    \includegraphics[width=\textwidth]{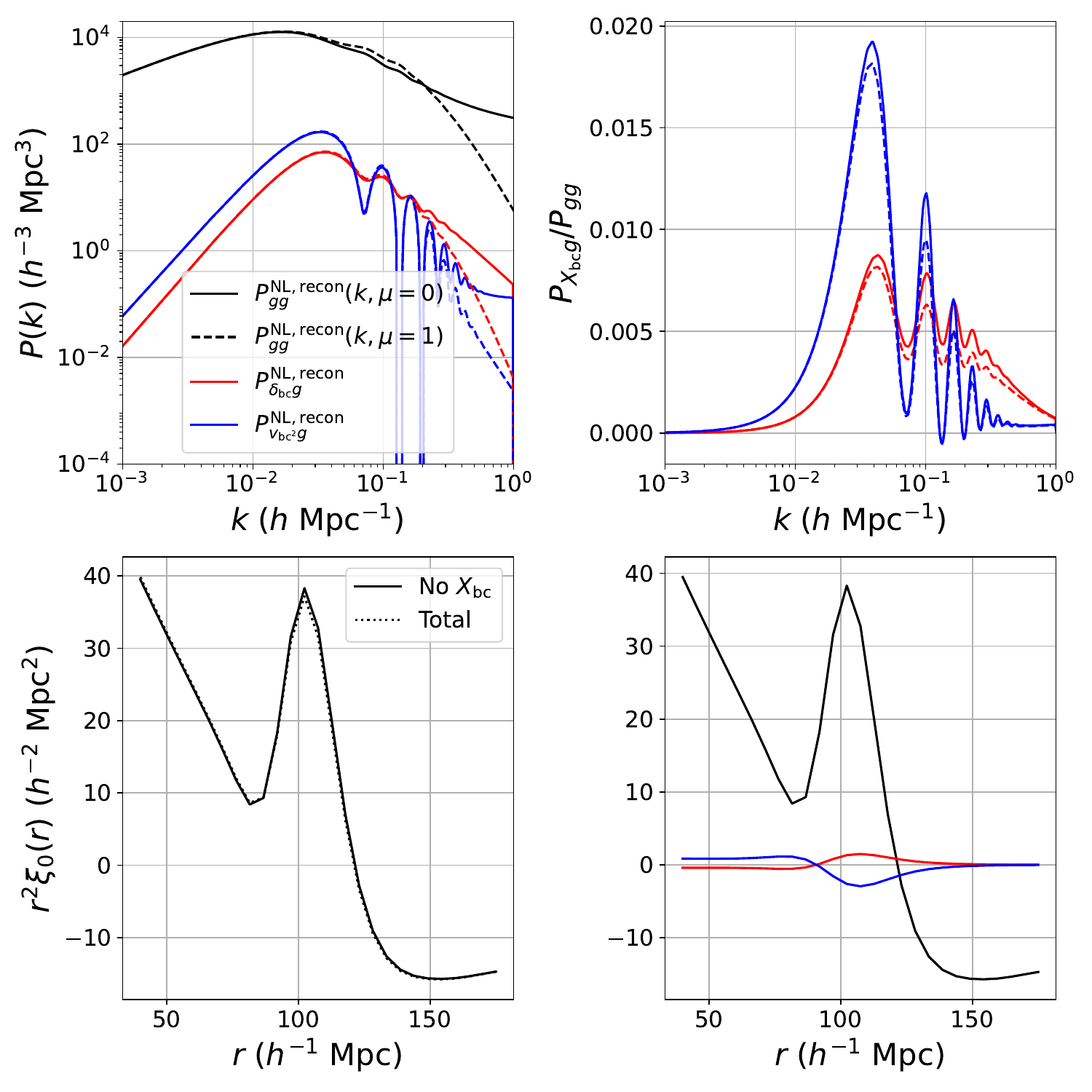}
    \caption{Contributions of relative density and velocity terms to the galaxy power spectrum. \textit{Top left:} Black shows the power spectrum without relative velocity and density terms; red shows the contribution from the galaxy-relatve density $\delta_{\textrm{bc}}$ term (fiducial value of $b_{\delta_{\textrm{bc}}} = 0.5$); and blue shows the contribution from the galaxy-streaming velocity disperson $v_{\textrm{bc}}^2$ term ($b_{v_{\textrm{bc}}^2} = 0.01 \sigma_{\textrm{bc}}^{-2}$, 10 times larger than the fiducial value for ease of display). Solid lines are for $\mu = 0$ and dashed lines for $\mu = 1$, using the approximate exponential model to relate the leading-order templates for $P_{\delta_{\textrm{bc}},\delta}$ and $P_{v_\textrm{bc}^2 \delta}$ to the post-reconstruction nonlinear power spectrum. \textit{Top right:} Fractional contribution of $P_{\delta_{\textrm{bc}},\delta}$ and $P_{v_\textrm{bc}^2 \delta}$
    to the total power spectrum.
    \textit{Bottom left:} Correlation function monopole for both the total power spectrum (dashed) and the galaxy density terms only (solid), using the fiducial values of $b_{\delta_{\textrm{bc}}} = 0.5$ and   $b_{v_{\textrm{bc}}^2} = 0.001 \sigma_{\textrm{bc}}^{-2}$. \textit{Bottom right:} Contributions of the $P_{\delta_{\textrm{bc}},\delta}$ and $P_{v_\textrm{bc}^2 \delta}$ terms ($b_{\delta_{\textrm{bc}}} = 0.5$ and $b_{v_{\textrm{bc}}^2} = 0.01 \sigma_{\textrm{bc}}^{-2}$).
    \label{fig:relative_velocity_power}}
\end{figure*}

The dominant contributions to the galaxy power spectrum are the cross-terms, $b_1 b_{\delta_{\textrm{bc}}} P_{\delta_{\textrm{bc}}\delta}(k)$ and $b_1 b_{v_{\textrm{bc}}^2} P_{v_\textrm{bc}^2 \delta}(k)$ \cite{KP4s2-Chen}. Because $\theta_{\textrm{bc}}$ is much smaller than $\delta_{\textrm{bc}}$, the $P_{\theta_{\textrm{bc}} \delta}$ term is suppressed.
The $P_{v_\textrm{bc}^2 \delta}$ is the relative velocity effect from \cite{TseliakhovichHirata}
arising from supersonic baryon-dark matter velocities. While it is a one-loop rather than linear order term,
it can be enhanced by the rms fluctuation of the relative velocities, $\sigma_{bc}$, $\sim$30 km s$^{-1}$ at recombination.

The relative velocity terms affect both the amplitude and position of the BAO peak.
Their impact on the BAO peak position
was shown to be small relative to the DESI DR1 errors in \cite{KP4s2-Chen}, and also small relative to the final expected errors from DESI.
We will perform a similar analysis to estimate the impact of the relative velocity errors on $\gamma_b$.


A full treatment of these terms at one-loop order in the post-reconstruction redshift-space power
spectrum is challenging (see \cite{Schmidt16,Chen19a} for the pre-reconstruction redshift space power spectrum in perturbation theory). We therefore proceed with an approximate treatment in which we add $P_{\delta_{\textrm{bc}},\delta}$ and $P_{v_\textrm{bc}^2 \delta}$
to a noiseless mock, fit the BAO amplitude in our baseline model, and measure the shift in $\gamma_b$.
We use an approximate damping model to propagate leading-order templates for $P_{\delta_{\textrm{bc}} \delta}$ (linear-order) and $P_{v_\textrm{bc}^2 \delta}$ (one-loop)
to the nonlinear, post-reconstruction field. We defer improvements in this modelling to future work in which we aim to jointly fit $\gamma_b$, $b_{\delta_{\textrm{bc}}}$, and $b_{v_{\textrm{bc}}^2}$.
In this approximate model, the contributions of $P_{\delta_{\textrm{bc}},\delta}$ and $P_{v_\textrm{bc}^2 \delta}$ take the form:
\begin{multline}
P_{\delta_{\textrm{bc}}\delta}^{\textrm{NL,recon}}(k, \mu) = 2 (b_1 + f\mu^2) \, b_{\delta_{\textrm{bc}}} \, P_{\delta_{\textrm{bc}}\delta}  (k,z=0) \, D(z) \, \times \\ \exp\bigg(-0.5 k^2 \left[ (1-\mu^2)\Sigma_{\perp}^2 + \mu^2 \Sigma_{\parallel}^2\right]\bigg) \left(\frac{1}{1 + 0.5 k^2 \mu^2 \Sigma_{\textrm{fog}}^2}\right)^2
\end{multline}
\begin{multline}
P_{v_{\textrm{bc}^2}\delta}^{\textrm{NL,recon}}(k, \mu) = (b_1 + f\mu^2) \, b_{v_{\textrm{bc}^2}} \, P_{v_{\textrm{bc}^2}\delta}  (k,z=0) \, D^2(z) \, \times \\ \exp\bigg(-0.5 k^2 \left[ (1-\mu^2)\Sigma_{\perp}^2 + \mu^2 \Sigma_{\parallel}^2\right]\bigg) \left(\frac{1}{1 + 0.5 k^2 \mu^2 \Sigma_{\textrm{fog}}^2}\right)^2
\end{multline}
where the linear and one-loop calculations of $P_{\delta_{\textrm{bc}} \delta}$ and $P_{v_\textrm{bc}^2 \delta}$ are from \cite{KP4s2-Chen}.
The scaling with the growth factor is different: for $P_{\delta_{\textrm{bc}},\delta}$,
$\delta$ grows with $D(z)$ but $\delta_{\textrm{bc}}$ is constant, causing the power spectrum to scale with only a single power of the growth factor.
For $P_{v_\textrm{bc}^2 \delta}$, $v_{\textrm{bc}}^2$ is constant in redshift but the leading contribution is at one-loop order, hence leading to two powers of the growth factor.
Rather than split $P_{\delta_{\textrm{bc}} \delta}$ and $P_{v_\textrm{bc}^2 \delta}$ into wiggle and no-wiggle components, we damp the entire power spectrum. This suppresses the amplitude of the no-wiggle component, but the dominant impact on the observed power spectrum comes from the oscillatory part, with $<1$\% contribution from the broadband. Our results are therefore minimally affected by the extra damping on the broadband component. In Fig.~\ref{fig:relative_velocity_power}, we show the contributions from $P_{\delta_{\textrm{bc}} \delta}$ and $P_{v_\textrm{bc}^2 \delta}$
to the galaxy power spectrum.

The impact of these terms depends on the unknown bias coefficients,
$b_{\delta_{\textrm{bc}}}$ and $b_{v_{\textrm{bc}}^2}$. For galaxies with a similar stellar mass to DESI and BOSS, \cite{Barreira20} measures $b_{\delta_{\textrm{bc}}}$
using the separate-universe technique in hydrodynamical simulations and finds $| b_{\delta_{\textrm{bc}}}| \sim 0.5$, with an increasing trend towards higher halo masses and redshifts. Similarly, \cite{Khoraminezhad20} finds $b_{\delta_{\textrm{bc}}} \sim -0.4$ for BOSS-like galaxies using separate-universe $N$-body simulations. The relative velocity bias is more poorly estimated, ranging between $10^{-5}$ $\sigma_{\textrm{bc}}^{-2}$  to $0.01 \sigma_{\textrm{bc}}^{-2}$ \cite{Yoo11,Blazek16,Schmidt16}. Unless galaxy formation preserves significant memory of early times, smaller values of $b_{v_{\textrm{bc}}^2}$ are more likely \cite{YooSeljak13}. We therefore assume a fiducial value of $b_{v_{\textrm{bc}}^2} = 10^{-3} \sigma_{\textrm{bc}}^{-2}$.

We measure $\gamma_b$ from the mock datasets with relative density and velocity perturbations using the post-reconstruction fits, and find 
\begin{equation}
\Delta \gamma_b = -0.0040 \, \left(\frac{b_{\delta_{\textrm{bc}}}}{0.5}\right)
\end{equation}
\begin{equation}
\Delta \gamma_b = -0.0015 \, \left(\frac{b_{v_{\textrm{bc}}^2}}{10^{-3}}\right)
\end{equation}
This is at most 0.40$\sigma$ for $b_{\delta_{\textrm{bc}}}$, justifying our 
decision to neglect these terms in our fits. However, for future datasets with higher precision, we will be required to marginalize over these terms. The sizes of these biases can be constrained by their impact on the power spectrum broadband in full-shape fits \cite{Blazek16,Beutler17,Slepian18} and indeed BOSS is already nearly large enough to constrain $b_{v_{\textrm{bc}}^2}$ \cite{Beutler17}.

\section{Comparison to previous baryon fraction fitting methodology}
\label{sec:compare_to_old_method}

In this section, we compare our baseline method, using \cite{Crespi25} to define $\gamma_b$ directly in \texttt{CAMB}, to our old fitting method used in the BOSS analysis \cite{BOSSBAOAmp}, in which we split the transfer function into baryonic and CDM components.

\subsection{EFT priors in the CLASS-PT based pipeline}

In \cite{BOSSBAOAmp}, we used a different EFT code (\texttt{CLASS-PT}), whereas we now use \texttt{Velocileptors}, to match the default code used in the DESI full-shape analysis \cite{DESI2024.V.KP5}. In order to consider the impact of this choice, we need to match the priors adopted for the \texttt{CLASS-PT} based analysis to those we now adopt using \texttt{Velocileptors}.

In the \texttt{CLASS-PT} bias basis (``East Coast'' basis), there are  11 free nuisance parameters per tracer (4 biases $b_1$, $b_2$, $b_{\mathcal{G}_2}$, $b_{\Gamma_3}$, 4 counterterms $c_0$, $c_2$, $c_4$, $\tilde{c}$, and 3 stochastic parameters $P_{\textrm{shot}}$, $a_0$, $a_2$).
The priors used in \cite{BOSSBAOAmp} are shown in Table~\ref{tab:classpt_priors}. These are 
very similar to the priors used in \cite{Philcox_Ivanov}, except that we halve the width of the priors on the counterterms and stochastic parameters to reduce prior volume effects.
We refer to these as the ``BOSS priors.''

\begin{table*}[]
    \centering
    \begin{tabular}{l|c|c}
    Parameter & BOSS prior & DESI prior, $b^n$ basis \\
    \hline
$b_1$ & flat[0,4]  & flat[0,4] \\
$b_2$ & $\mathcal{N}(0,1^2)$ & $\mathcal{N}(0,5^2)$ \\
$b_{\mathcal{G}_2}$ & $\mathcal{N}(0,1^2)$ & $\mathcal{N}(0,5^2)$ \\
$b_{\Gamma_3}$ & $\mathcal{N}\left(\frac{23}{42}(b_1-1),1^2\right)$ & $\frac{23}{42}(b_1-1)$ \\
$c_0$ [Mpc/$h$]$^2$ & $\mathcal{N}(0,15^2)$ & $\mathcal{N}(0, \sigma_{c_0}^2$), $\sigma_{c_0}$ = [19.1, 25.5, 33.9, 34.5, 17.0, 35.2] \\
$c_2$ [Mpc/$h$]$^2$ & $\mathcal{N}(30,15^2)$ & $\mathcal{N}(0, \sigma_{c_0,i}^2$), $\sigma_{c_0}$ = [21.4, 24.7, 28.7, 28.9, 19.9, 29.2] \\
$c_4$ [Mpc/$h$]$^2$ & $\mathcal{N}(0,15^2)$ & 0 \\
$\tilde{c}$ [Mpc/$h$]$^4$ & $\mathcal{N}(500,250^2)$  & 0\\
$P_{\textrm{shot}}$ & $\mathcal{N}(0,0.5^2)$ & $\mathcal{N}(0,2^2)$ \\
$a_0$ & $\mathcal{N}(0,0.5^2)$ & 0 \\
$a_2$ & $\mathcal{N}(0,0.5^2)$ & $\mathcal{N}(0,\sigma_{a_2,i}^2)$, $\sigma_{a_2}$ = [6.2, 6.2, 6.2, 6.2, 0.85, 0.99] \\

    \end{tabular} 
    \caption{Priors on Eulerian bias parameters, counterterms, and stochastic terms used when measuring $\gamma_b$ using the CLASS-PT pipeline used in \cite{BOSSBAOAmp}. The prior used in that work (``BOSS prior'') is shown in the left column. The right column shows the best-validated prior for the DESI analysis using this pipeline. The widths of these priors are motivated to match the priors used in the DESI Full-Shape analysis \cite{DESI2024.V.KP5}, but we sample in the bias parameters rather than re-scaling them by powers of $\sigma_8$ to reduce projection effects in $\gamma_b$.
    \label{tab:classpt_priors}}
\end{table*}

While these priors were sufficient for recovering unbiased cosmology on BOSS mocks and noiseless theory vectors using the BOSS covariance, we find that the prior volume effects are more severe on DESI data and these priors are no longer sufficient. To recover unbiased cosmology with the ``old'' pipeline, we therefore use a set of priors derived from translating the DESI priors to the East Coast basis. These priors are not identical to the DESI priors, because they are defined in a different basis so the volume measure is different, but we instead match the prior widths on each parameter individually.

The DESI priors are defined in terms of Lagrangian bias parameters. These can be translated to the Eulerian bias parameters used by \texttt{CLASS-PT}:
\begin{equation}
    b_2^E = b_2^L + \frac{8}{21} b_1^L
\end{equation}
\begin{equation}
    b_{\mathcal{G}_2}^E = b_s^L - \frac{2}{7} b_1^L
\end{equation}
In the DESI full-shape analysis, the following priors are applied to the bias parameters:
\begin{equation}
(1 + b_1^L)\sigma_8(z) \rightarrow U(0,3)
\end{equation}
\begin{equation}
    b_2^L \sigma_8(z)^2 \rightarrow N(0,5^2)
\end{equation}
\begin{equation}
    b_s^L \sigma_8(z)^2 \rightarrow N(0,5^2)
\end{equation}
We can then express the prior on $b_2^E$, $\sigma_{b_2^E}$ using error propagation:
\begin{equation}
    \sigma^2_{b_2^E \sigma_8(z)^2} = \sigma^2_{b_2^L \sigma_8(z)^2} + \left(\frac{8}{21}\right)^2 \sigma_8(z)^2 \sigma^2_{b_1^L \sigma_8(z)}
\end{equation}
The standard deviation of the uniform distribution between 0 and 3 is 0.87, while the standard deviation of $b_2^L \sigma_8(z)^2$ is 5. Hence the second term is $\sim$0.3\% of the first term at $z = 0$ and smaller at higher redshift. Likewise, the second term only shifts the mean of $b_2^E \sigma_8(z)^2$ by $\sim$0.2 at $z=0$, tiny compared to the width of the prior. Hence the prior on $b_2^E$ can be approximated by:
\begin{equation}
b_2^E \sigma_8(z)^2 \rightarrow N(0,5^2)
\end{equation}
Similarly, for $b_{\mathcal{G}_2}^E$, the second term is $\sim$0.2\% of the first term, and the prior is also the same as on $b_s^L$:
\begin{equation}
b_{\mathcal{G}_2}^E \sigma_8(z)^2 \rightarrow N(0,5^2)
\end{equation}
Finally, in the DESI prior basis, the third-order bias is set to zero, hence we fix $b_{\Gamma_3}$ to zero.

\begin{figure}
    \includegraphics[width=0.45\textwidth]{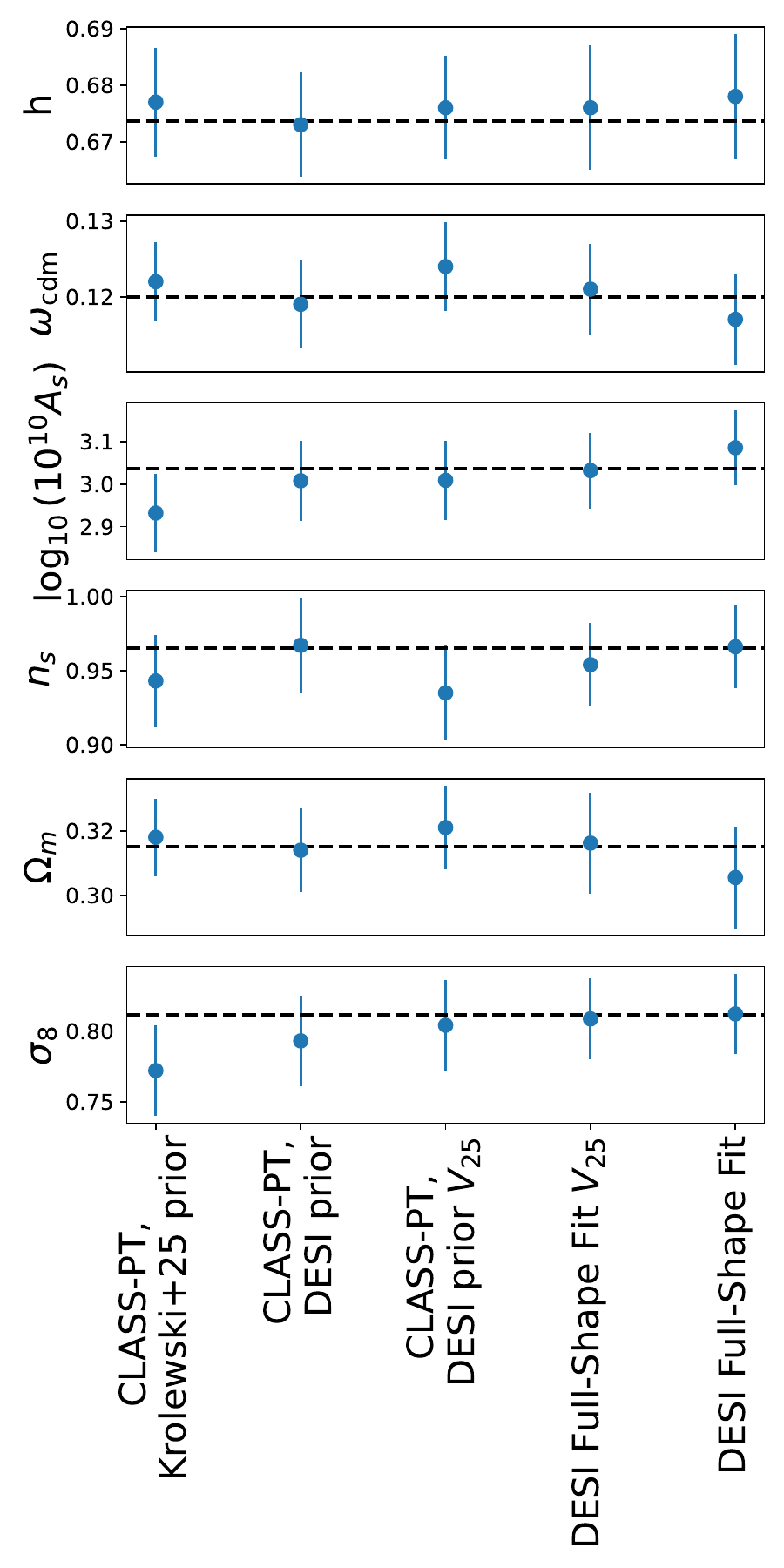}
    \caption{Comparison of $\Lambda$CDM constraints between our pipeline and the DESI Full-Shape pipeline used in \cite{DESI2024.V.KP5}. The left-most set of points comes from CLASS-PT run with the prior from Krolewski+25 \cite{BOSSBAOAmp} (very similar to the default CLASS-PT priors in \cite{Philcox_Ivanov}), fitting the mean of the 25 Abacus mocks and using the covariance matrix from the DESI data. The next set of points also uses CLASS-PT, but with a bias prior matching the choices of \cite{DESI2024.V.KP5}. The third set uses the same fitting method as the second set, but scaling the covariance by a factor of 25 to appropriately match the fluctuations in the 25$\times$ larger volume considered. The fourth and fifth points are taken from the DESI fits using Velocileptors and the nuisance parameter priors from Table 4 in \cite{DESI2024.V.KP5}, either using the DESI Y1 covariances (fifth column) or the covariance scaled down by 25 (fourth column).
\label{fig:classpt_velocileptors_consistency}
    }
\end{figure}

\begin{figure*}
    \includegraphics[width=\textwidth]{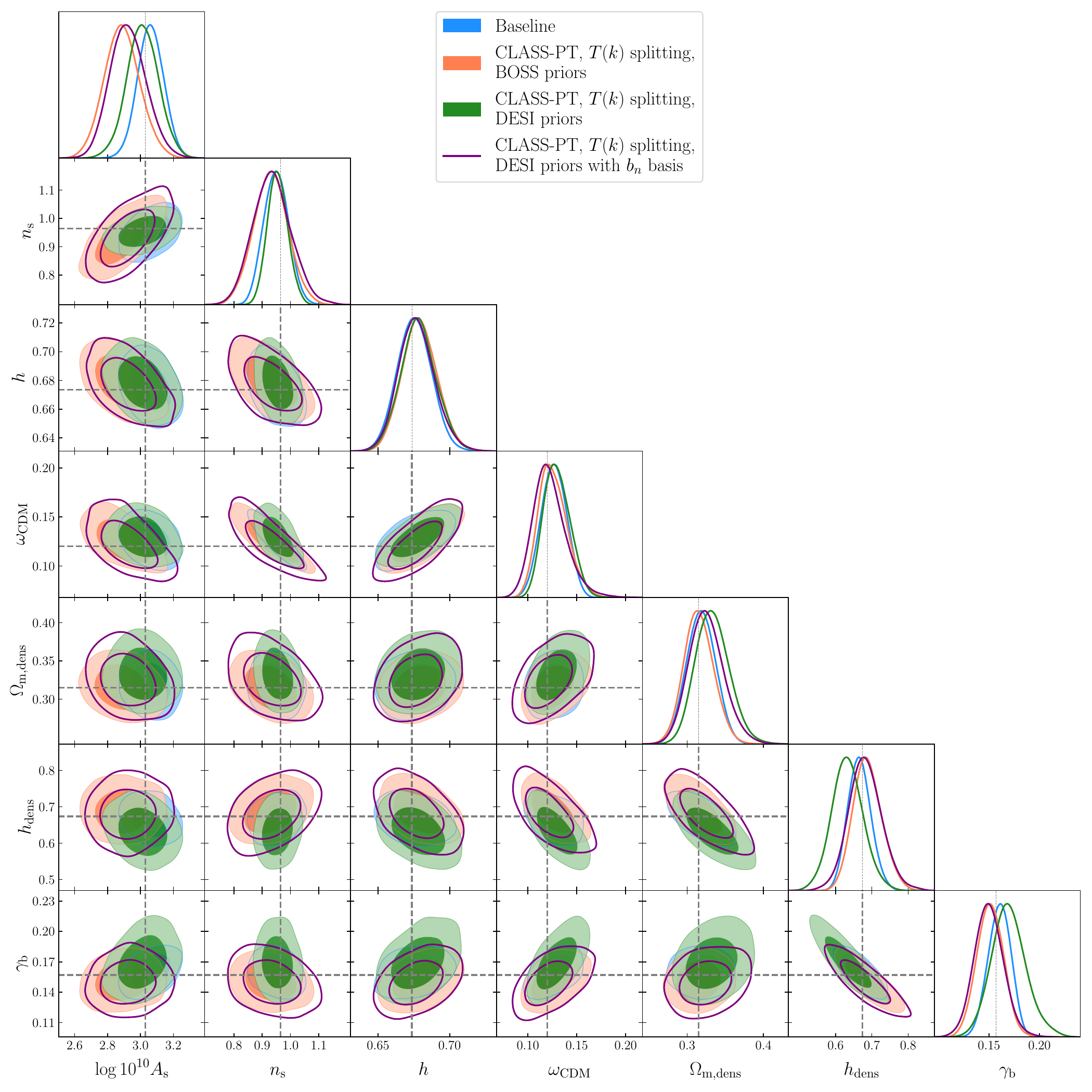}
    \caption{Comparison of our results on Abacus mocks using different splitting methods and bias bases. The blue contours show the baseline results presented in the main body of the paper: \texttt{Velocileptors} EPT bias parameters with the HOD-informed prior to migitate projection effects, using the consistent splitting of \cite{Crespi25}.
    The orange contours show our old CLASS-PT pipeline used in \cite{BOSSBAOAmp} applied to the DESI mocks. The green contours show this same pipeline, but using the DESI bias priors, leading to large prior volume effects on $\gamma_b$ and $h^{\textrm{dens}}$. The purple contours show the CLASS-PT pipeline with the DESI bias priors, but without rescaling the $n^{\textrm{th}}$ order biases by $\sigma_8^n$. This produces very similar constraints to the bias prior used in \cite{BOSSBAOAmp}, but mitigates projection effects on noiseless mocks.
    \label{fig:k25_vs_d25_priors}}
\end{figure*}

The counterterm priors are set by specifying that the counterterm contribution to the power spectrum is at most 0.5$\times$ linear theory. The CLASS-PT counterterm contribution is given by:
\begin{align}
    P_{\textrm{ctr}}^{\textrm{CLASS-PT}} = -2 \tilde{c_0}k^2 P_{\textrm{lin}}(k) - 2 \tilde{c_2} f \mu^2 k^2 P_{\textrm{lin}}(k) - \nonumber \\ 2 \tilde{c_4} f^2 \mu^4 k^2 P_{\textrm{lin}}(k) - \tilde{c}(z)f^4 \mu^4 k^4 (b_1 + f\mu^2)^2 P_{\textrm{lin}}(k)
\label{eqn:pctr}
\end{align}
where the basis $\tilde{c_0}$, $\tilde{c_2}$, $\tilde{c_4}$ is rotated into the basis $c_0$, $c_2$, $c_4$ on which the prior is defined:
\begin{equation}
    c_0 = \tilde{c_0} + \frac{f}{3} \tilde{c_2} + \frac{f^2}{5} \tilde{c_4}
\end{equation}
\begin{equation}
    c_2 = \tilde{c_2} + \frac{6 f}{7} \tilde{c_4}
\end{equation}
\begin{equation}
    c_4 = \tilde{c_4}
\end{equation}
The linear contribution to the power spectrum is:
\begin{equation}
    P_{\textrm{lin cont.}} = (b_1^E + f\mu^2)^2 P_{\textrm{lin}}(k)
\label{eqn:plin}
\end{equation}
Since we are defining prior widths, we can ignore the overall negative sign in Eq.~\ref{eqn:pctr}, and then by 
setting Eq.~\ref{eqn:pctr} and $0.5\times$ Eq.~\ref{eqn:plin} equal order-by-order in $\mu$, we obtain:
\begin{equation}
    2 \tilde{c_0}k^2 = 0.5 (b_1^E)^2
\end{equation}
\begin{equation}
    2 \tilde{c_2}k^2 = b_1^E
\end{equation}
Since we do not use the hexadecapole, we fix $c_4$ to zero (as in the DESI full-shape prior).
We can then compute prior widths on $c_0$ and $c_2$, adding the two terms in $c_0$ in quadrature since they are independent. We use
the best-fit $b_1^E$ from the DESI full-shape fits to Abacus and evaluate at the effective redshifts of the Abacus mocks (0.2, 0.5, 0.8, 0.8, 1.325, 1.4):
\begin{equation}
\begin{split}
&\frac{c_{0,i}}{(\textrm{Mpc/$h$})^2} \sim \mathcal{N}(0, \sigma_{c_{0},i} ^2) \\
&\sigma_{c_0} = [19.1, 25.5, 33.9, 34.5, 17.0, 35.2]
\end{split}
\end{equation}
\begin{equation}
\begin{split}
&\frac{c_{2,i}}{(\textrm{Mpc/$h$})^2} \sim \mathcal{N}(0, \sigma_{c_{2},i} ^2) \\
&\sigma_{c_2} = [21.4, 24.7, 28.7, 28.9, 19.9, 29.2]
\end{split}
\end{equation}
Since the $\tilde{c}$ counterterm only affects the hexadecapole, it is also set to zero.

Finally, we translate the shot noise parameters, where the contributions to the stochastic terms are
\begin{equation}
    P_{\textrm{sto}}^{\textrm{CLASS-PT}} = \frac{1}{\bar{n}} \left(P_{\textrm{shot}}  + a_0 \left(\frac{k}{k_{\textrm{NL}}}\right)^2 + a_2 \mu^2 \left(\frac{k}{k_{\textrm{NL}}}\right)^2\right)
\end{equation}
and in \texttt{Velocileptors}
\begin{equation}
    P_{\textrm{sto}}^{\textrm{vel.\ }} = \frac{1}{\bar{n}} \left(\textrm{SN}_0 + \textrm{SN}_2 \mu^2 k^2\right)
\end{equation}
Hence, the prior on $P_{\textrm{shot}}$ is the same as that on $\textrm{SN}_0$:
\begin{equation}
P_{\textrm{shot}} \sim \mathcal{N}(0, 2^2)
\end{equation}
and $a_0$ is fixed to zero since there is no $\propto \mu^0 k^2$ stochastic term in \texttt{Velocileptors}.
The prior on $\textrm{SN}_2$ is
\begin{equation}
    \textrm{SN}_2 \rightarrow N(0,5^2) \times f_{\textrm{sat}} \frac{\sigma_{\textrm{eff}}^2}{\bar{n}_g}
\end{equation}
Using the satellite fractions and velocity dispersions
from Table 2 in \cite{KP5s2-Maus}, we find
\begin{equation}
\begin{split}
&a_{2,i} \sim \mathcal{N}(0, \sigma_{a_{2,i}}^2) \\
&\sigma_{a_2} = [6.2, 6.2, 6.2, 6.2, 0.85, 0.99]
\end{split}
\end{equation}

When applying these priors to $\Lambda$CDM constraints, we show in Fig.~\ref{fig:classpt_velocileptors_consistency} that they lead to good consistency between our pipeline and the fiducial DESI full-shape results on Abacus mocks, with some minor differences in $n_s$ and $\omega_{\textrm{cdm}}$ likely due to slight differences in perturbative expansions. Most notably, using the DESI bias basis and prior substantially mitigates the projection effects in $\sigma_8$, reducing the $\sim$1$\sigma$ bias on $\sigma_8$ with the CLASS-PT priors. The remaining differences in $\sigma_8$ disappear when the covariance is scaled down by a factor of 25 (approaching the maximum of the posterior in both the CLASS-PT and DESI runs).

\subsection{Results on mocks}

\begin{figure*}
    \includegraphics[width=\textwidth]{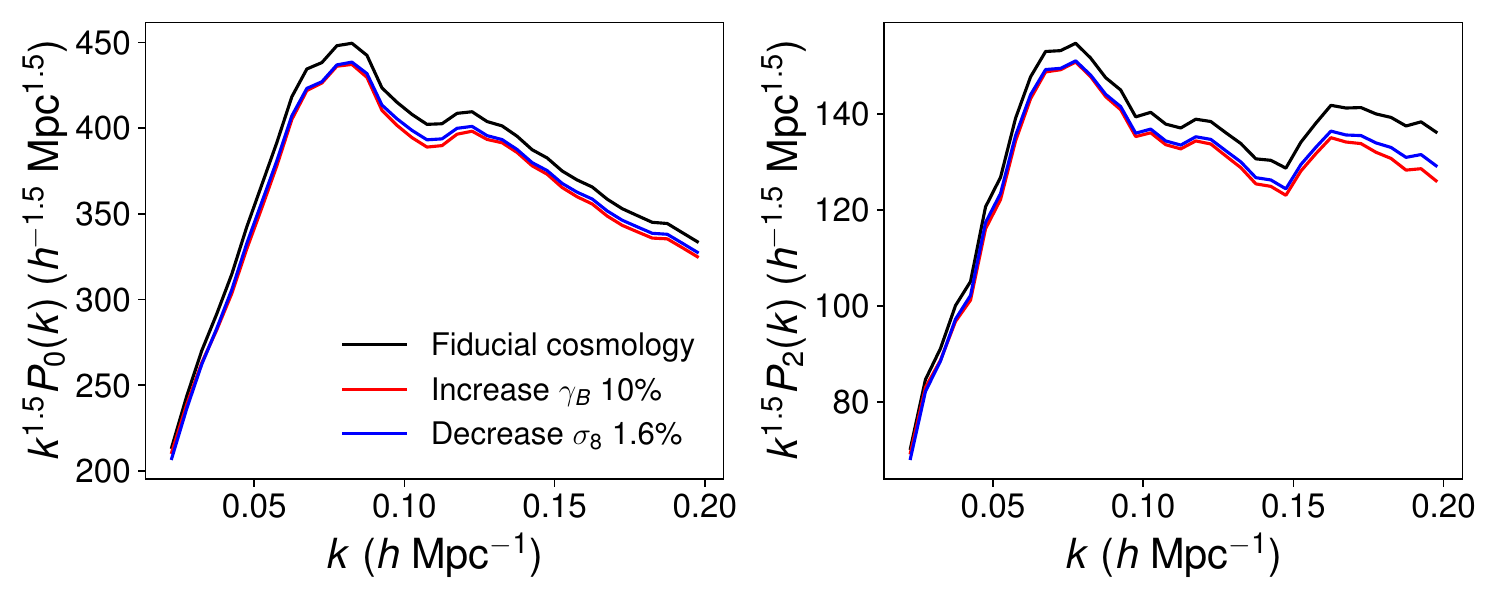}
    \caption{Illustration of the $\gamma_b$-$\sigma_8$ degeneracy. Theory predictions for the monopole and quadrupole in the fiducial cosmology (black), a cosmology with 10\% higher $\gamma_b$ (red), and a cosmology with 1.6\% lower $\sigma_8$ (blue).
    Due to the baryons' effect on the galaxy power spectrum shape, increasing $\gamma_b$ suppresses the power spectrum on scales smaller than the baryonic Jeans scale. The degeneracy is broken by the more prominent BAO wiggles in models with larger $\gamma_b$.
    \label{fig:sigma8_gammaB_degeneracy}}
\end{figure*}

In Fig.~\ref{fig:k25_vs_d25_priors}, we compare parameters inferred from the Abacus mocks from our baseline pipeline from the main text of the paper to the CLASS-PT-based pipeline used in the previous analysis.
We find that the DESI priors (defined in the $b_n \sigma_8^n$ basis) have large projection effects on $h^{\textrm{dens}}$ and $\gamma_b$.
These prior volume effects arise from the degeneracy between $\gamma_b$ and $\sigma_8$. The degeneracy is shown in Fig.~\ref{fig:sigma8_gammaB_degeneracy}, which compares the impact of changing $\sigma_8$ and $\gamma_b$ on the model monopole and quadrupole. This arises from the suppression of the power spectrum on scales smaller than the baryonic Jeans scale. This degeneracy means that in the basis where the bias parameters are scaled by $\sigma_8$, we make the rescaled bias parameters much more degenerate with $\gamma_b$, leading to dramatically worse prior volume effects on $\gamma_b$. 
We resolve this by using the DESI priors, but sampling in $b_n$, rather than $b_n \sigma_8^n$.

Apart from the DESI priors defined in the $b_n \sigma_8^n$ basis, the other three configurations produce consistent and unbiased constraints on $h^{\textrm{dens}}$ and $\gamma_b$. In particular, the results using CLASS-PT and the priors from our BOSS analysis \cite{BOSSBAOAmp} are very similar to those using CLASS-PT and the DESI matched priors with the $b_n$ basis.
We also summarize these results in Table~\ref{tab:full_shape_summary_appendix}, which shows that for the CLASS-PT pipeline, the BOSS priors have worse projection effects on noiseless $\Lambda$CDM mocks than the DESI priors in the $b_n$ basis. As a result, our default configuration in the CLASS-PT pipeline (which we also test on data) uses the DESI priors in the $b_n$ basis.

The baseline constraints from the updated splitting method of \cite{Crespi25} have slightly tighter constraints on $\gamma_b$ and $h_{\textrm{dens}}$ than the old $T(k)$ splitting method, since the new method essentially allows the baryon fraction to affect the baryon and CDM transfer functions. This allows the new method to consistently extract all of the baryon fraction information.

We also test the CLASS-PT pipeline (with the fiducial prior choice: DESI-matched priors with $b_n$ basis) on noiseless mocks in $\Lambda$CDM and EDE with different true values of the baryon fraction.
 We find generally good recovery of the input parameters, with biases on $f_b$ and $h$ typically $<0.5\sigma$
and similar performance on the $\Lambda$CDM and EDE mocks. These biases are somewhat larger than the 0.1-0.2$\sigma$ biases found in our baseline method.

\begin{table*}[]
    \centering
    \begin{tabular}{l|cc|cc|cc}
    & \multicolumn{2}{c|}{Truth} & \multicolumn{2}{c|}{$h$} &  \multicolumn{2}{c}{$\gamma_b$} \\
    Abacus run & $h$ & $f_b$ & Mean $\pm$ 1$\sigma$ (MAP) & $n\sigma$ &  Mean $\pm$ 1$\sigma$ (MAP) & $n\sigma$ \\
    \hline
     Baseline & 0.6736 & 0.1571 & $0.665 \pm 0.031$ (0.662) & -0.28 & $0.160 \pm 0.011$ (0.160) & 0.26 \\
    Baseline, HIP, $T(k)$ splitting & 0.6736 & 0.1571 & $0.662 \pm 0.030$ (0.663) & -0.39 & $0.163 \pm 0.011$ (0.160) & 0.54 \\
    Baseline, no HIP, $T(k)$ splitting & 0.6736 & 0.1571 & $0.672 \pm 0.031$ (0.663) & -0.05 & $0.157 \pm 0.011$ (0.160) & -0.01 \\
    CLASS-PT, BOSS priors & 0.6736 & 0.1571 & $0.681^{+0.042}_{-0.037}$ (0.676) & 0.21 & 
    $0.153^{+0.012}_{-0.013}$ (0.157) & -0.34 \\
    CLASS-PT, DESI priors $b_n$ basis & 0.6736 & 0.1571 & $0.680^{+0.047}_{-0.043}$ (0.689) & 0.16 & $0.150^{+0.015}_{-0.014}$ (0.150) & -0.52 \\
     \hline
    $\Lambda$CDM noiseless mocks \\
     \hline
    Baseline & 0.6736 & 0.1571 & $0.681 \pm 0.037$ (0.677) & 0.20 & $0.155 \pm 0.0144$ (0.155)  & -0.14 \\
    CLASS-PT, BOSS priors & 0.6736 & 0.1571 & $0.700^{+0.042}_{-0.040}$ (0.674) & 0.66 & $0.146^{+0.013}_{-0.012}$ (0.157) & -0.85 \\
    CLASS-PT, DESI priors $b_n$ basis & 0.6736 & 0.1571 & $0.681 \pm 0.037$ (0.674) & 0.20 & $0.155 \pm 0.014$ (0.157)  & -0.14 \\
    \hline
    EDE noiseless mocks \\
    \hline
    Baseline & 0.7219 & 0.1471 & $0.721 \pm 0.040$ (0.716)  & -0.03 & $0.1479 \pm 0.015$ (0.150) & 0.05 \\
    DESI priors $b_n$ basis  & 0.7219 & 0.1471 & 
     $0.706^{+0.046}_{-0.042}$ & 0.35 & $0.144^{+0.013}_{-0.013}$ & -0.21 \\
    \end{tabular} 
    \caption{Summary of full-shape fits to the mean of the 25 Abacus mocks and noiseless theory vectors in both the default $\Lambda$CDM and early dark energy cosmologies. For the Abacus mocks, we compare results from different fitting pipelines, changing from the baseline analysis of this paper to the
    pipeline used in \cite{BOSSBAOAmp}. The top row shows the baseline approach, using \texttt{Effort.jl}, the consistent \texttt{CAMB} approach of \cite{Crespi25} to model $\gamma_b$ and the HOD-informed prior (HIP) to minimize prior volume effects.
    The next row shows the baseline approach, but instead using the old transfer function ($T(k)$) splitting from the BOSS analysis.
    The third row also omits the HOD-informed prior.
    The fourth and fifth rows in the Abacus runs compare the CLASS-PT pipeline, both with the priors used in the BOSS analysis and the priors matched to DESI (but constraining the bias parameters $b_n$ rather than the combination $b_n \sigma_8^n$).
    The bottom rows show results on noiseless mocks, demonstrating that the BOSS priors used with CLASS-PT give large projection effects and therefore justifying our choice of the DESI priors with the $b_n$ basis for the CLASS-PT pipeline.
    \label{tab:full_shape_summary_appendix}}
\end{table*}

\subsection{Results on data}

\begin{figure*}
    \includegraphics[width=\textwidth]{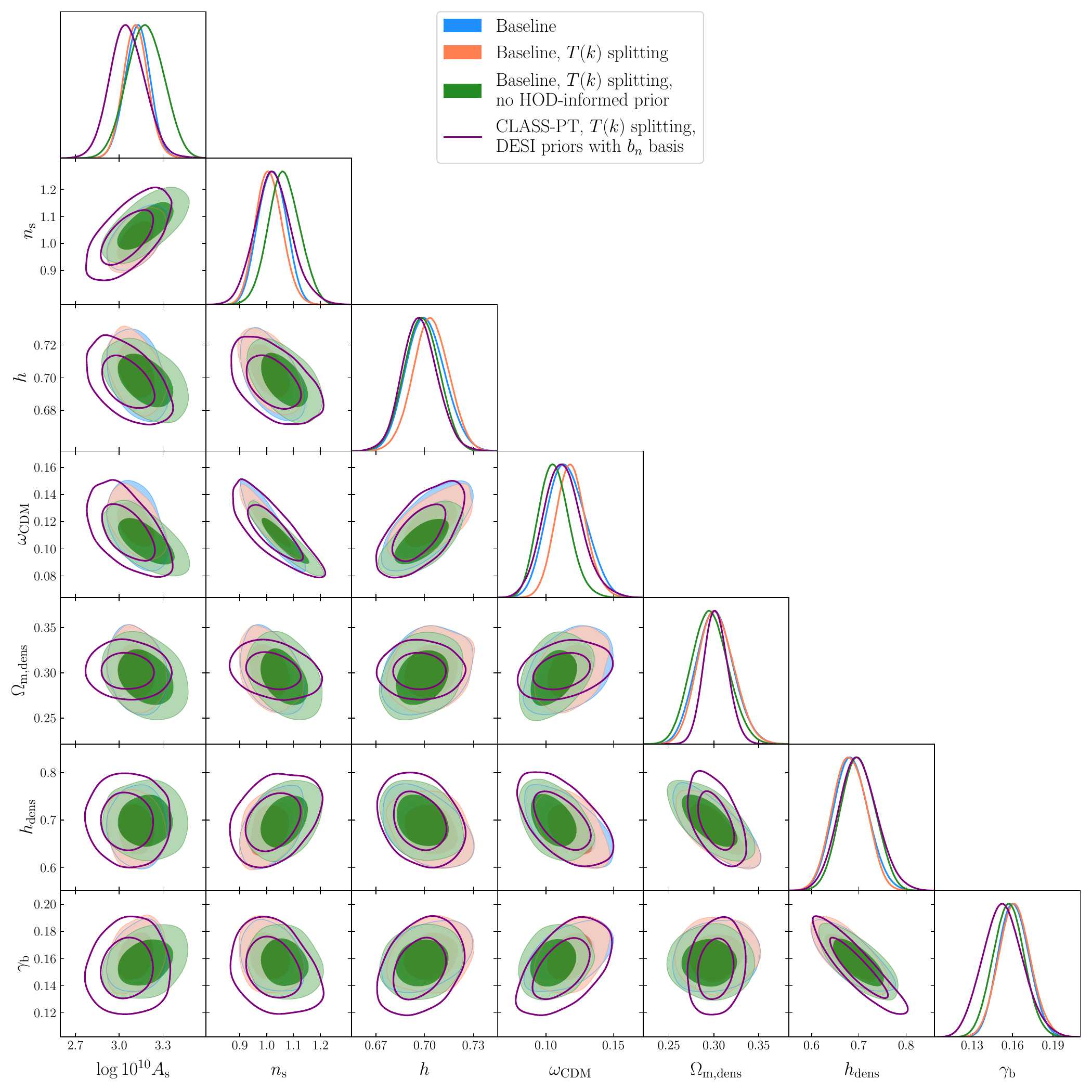}
    \caption{Comparison of energy-density fits to DESI data using different fitting pipelines. The blue contours are the baseline full-shape fit results, using \texttt{Effort.jl}, the new approach of \cite{Crespi25} for $\gamma_b$, and the HOD-informed prior. The orange and green contours show the same pipeline, but using the old transfer function based splitting (orange) or using the transfer function splitting and no HOD informed prior (green). The purple contours show the CLASS-PT pipeline used in \cite{BOSSBAOAmp}, using the modified DESI prior basis.
    \label{fig:data_compare_different_configurations}}
\end{figure*}

In Fig.~\ref{fig:data_compare_different_configurations} and Table~\ref{tab:data_results_compare_different_configuration}, we show that our
results on data are robust to the method used to measure the baryon fraction from the full shape of the power spectrum.
First, we find that variations in our baseline result do not change $H_0$ much; if we use the transfer function split to measure $\gamma_b$ rather than the new approach of \cite{Crespi25}, $H_0$ drops by 0.3 km s$^{-1}$ Mpc$^{-1}$.
If we also remove the HOD-informed prior, $H_0$ rises by 1.8 km s$^{-1}$ Mpc$^{-1}$.
If we instead use the CLASS-PT pipeline with DESI priors (in the $b_n$ basis where projection effects are minimized), we find a slightly higher $H_0$, with a slightly larger errorbar, than the baseline results. 
Since this run uses Ly$\alpha$ BAO, we compare it to our baseline run including Ly$\alpha$ and find good consistency in $\Omega_{\mathrm{m}}$ with a small offset in $\gamma_b$.
All of these differences are consistent with noise fluctuations given the slightly different statistical power of each result. Moreover, the changes are at most 0.3$\sigma$, well within our statistical uncertainty on $H_0$.


\begin{table*}[]
    \centering
    \begin{tabular}{l|ccc}
    Data combination & $H_0^{\textrm{dens}}$ (km s$^{-}$ Mpc$^{-1}$) & $\Omega_{\textrm{m,dens}}$ & $\gamma_b$ \\
    \hline
   Baseline (DESI DR1 galaxy clustering) & $68.4 \pm 3.6$ & $0.301 \pm 0.020$ & $0.161 \pm 0.010$  \\
    Baseline, HIP, $T(k)$ splitting & $68.1 \pm 3.4$ & $0.302 \pm 0.019$ & $0.162 \pm 0.012$ \\
    Baseline, no HIP, $T(k)$ splitting & $69.9 \pm 3.5$ & $0.296 \pm 0.020$  & $0.157 \pm 0.011$ \\
   Baseline (DESI DR1 galaxy clustering + Ly$\alpha$) & $68.6 \pm 3.1$ & $0.301 \pm 0.014$ & $0.161 \pm 0.010$  \\
    CLASS-PT, DESI priors $b_n$ basis  & \multirow{2}{*}{$69.7^{+4.1}_{-3.7}$}  & \multirow{2}{*}{$0.302^{+0.014}_{-0.013}$} & \multirow{2}{*}{$0.153 \pm 0.014$}  \\
    (DESI DR1 galaxy clustering + Ly$\alpha$) \\
    \end{tabular} 
    \caption{Summary of full-shape fits to the DESI data for the baseline results (top) and different analysis choices, either using the CLASS-PT pipeline, or using the \texttt{Effort.jl} pipeline but with the transfer function split or no HOD-informed prior. The tests of the baseline pipeline were done using DESI DR1 galaxy clustering only, while the CLASS-PT pipeline run also includes Ly$\alpha$ BAO, so we compare it to the baseline pipeline run with Ly$\alpha$ BAO for consistency.
    \label{tab:data_results_compare_different_configuration}}
\end{table*}

\section{Detailed comparison of post-reconstruction fits to DESI baseline}
\label{sec:post_recon_compare}

In Table~\ref{tab:abacus_post_recon_tests_appendix}, we 
compare the median of the 25 fits to the Abacus altmtl mocks between our pipeline to fit the BAO amplitude and the DESI baseline.
Our pipeline has similar constraining power on $\alpha_{\textrm{iso}}$ and slightly worse constraining power on $\alpha_{\textrm{AP}}$; this is due to the extra free parameters and broader priors on the damping parameters in our pipeline vs.\ the DESI baseline.

In Table~\ref{tab:detailed_comparison_alphas}, we compare in more detail the offset between the two pipelines for LRG at $0.4 < z < 0.6$. The first three rows reproduce the columns of Table~\ref{tab:abacus_post_recon_tests_appendix}, while the next rows show various other changes in the pipeline.
The dominant changes driving the offset in $\alpha_{\textrm{iso}}$ ($\Delta \alpha_{\textrm{iso}} = 0.008$) are the larger fitting range in our fits ($\Delta \alpha_{\textrm{iso}} = 0.002$) and
the fact that the baseline DESI fits sample in $\alpha_{\textrm{iso}}$ and $\alpha_{\textrm{AP}}$ 
rather than $\alpha_{\parallel}$ and $\alpha_{\perp}$.
The larger errors on $\alpha_{\textrm{AP}}$ are primarily driven by the extra freedom in our model,
with broader priors on the damping parameters, extra $\alpha$ parameters changing the broadband,
and varying $\gamma_b$ and $b_\partial$. These choices tighten $\alpha_{\textrm{AP}}$ constraints by $\sim$25\%. Parameterizing the bias with $b_1$ and $f$, as in the DESI baseline (rather than $B_0$ and $B_2$ in our fits) further decreases the $\alpha_{\textrm{AP}}$ error by 5\%. Offsets in $\alpha_{\textrm{AP}}$ are primarily driven by the nuisance parameter priors, and the basis to sample the $\alpha$ parameters, although there are also small shifts from the bias parameter set and the choice of whether to rescale the damping $C(k,\mu)$ by $\alpha$ (the baseline DESI analysis rescales $C(k,\mu)$ by $\alpha$, whereas we rescale by $\alpha_{\textrm{BB}}$ instead).

\begin{table*}[]
    \centering
    \begin{tabular}{l|c|c|c}
    Parameter  & Default, $\gamma_b$ & $\gamma_b$, $50 < r < 150$ $h^{-1}$ Mpc & DESI \\
    \hline
    \hline
    BGS, $0.1 < z < 0.4$ & & & \\
    \hline
    $\alpha_{\textrm{iso}}$ & $0.9990 \pm 0.0226$ & $0.9967 \pm 0.0194$ & $0.994 \pm 0.020$ \\
    \hline
    LRG, $0.4 < z < 0.6$ & & &  \\
    \hline
    $\alpha_{\textrm{iso}}$ & $1.0061 \pm 0.0155$ & $1.0043 \pm 0.0154$ & $0.9978 \pm 0.0135$ \\
    
    $\alpha_{\textrm{AP}}$ & $1.0084 \pm 0.0547$ &  $1.0087 \pm 0.0642$ & $0.9897 \pm 0.0478$ \\ 
    \hline
    LRG, $0.6 < z < 0.8$ & & & \\
    \hline
    $\alpha_{\textrm{iso}}$ & $0.9995 \pm 0.011$ & $0.9991 \pm 0.0114$ & $0.9946 \pm  0.011$\\
    $\alpha_{\textrm{AP}}$ &  $1.0106 \pm 0.0392$ & $1.0145 \pm 0.0403$ & $1.0081 \pm 0.036$ \\
    \hline
    LRG+ELG, $0.8 < z < 1.1$ & & &  \\
    \hline
    $\alpha_{\textrm{iso}}$ & $1.0038 \pm 0.0085$ & $1.0015 \pm 0.0085$ & $0.9990 \pm 0.0084$ \\
    $\alpha_{\textrm{AP}}$ &    $0.9960 \pm 0.0277$ & $1.0011 \pm 0.0290$ & $0.9930 \pm 0.0352$ \\
    \hline
    ELG, $1.1 < z < 1.6$ & & & \\
    \hline
    $\alpha_{\textrm{iso}}$ & $1.0032 \pm 0.0160$ & $0.9980 \pm 0.0157$ & $0.9949 \pm 0.015$ \\
    $\alpha_{\textrm{AP}}$ & $1.0037 \pm 0.0555$ & $1.0104 \pm 0.0575$ & $0.9986 \pm 0.051$ \\
    \hline
    QSO, $0.8 < z < 2.1$ & & & \\
    \hline
    $\alpha_{\textrm{iso}}$ & $0.9928 \pm 0.0179$ & $0.9991 \pm 0.0172$ & $0.995 \pm 0.018$ \\
    \end{tabular}
    \caption{Comparison of BAO scaling parameters (the median of the 25 Abacus altmtml mocks, and the median of their errors) in our pipeline with its default settings (left); our pipeline matching the DESI fitting range $50 < r < 150$ $h^{-1}$ Mpc (center); and the default DESI pipeline from Table 9 in \cite{DESI.DR2.BAO.cosmo}. Unlike in Table~\ref{tab:abacus_post_recon_tests}, all $\alpha$ parameters are relative to the DESI fiducial cosmology.
    \label{tab:abacus_post_recon_tests_appendix}}
\end{table*}

\begin{table}[]
    \centering
    \begin{tabular}{l|cccc}
    Pipeline & $\alpha_{\textrm{iso}}$ & $\sigma_{\alpha_{\textrm{iso}}}$ & $\alpha_{\textrm{AP}}$ & $\sigma_{\alpha_{\textrm{AP}}}$  \\
    \hline
    DESI & 0.998 & 0.014 & 0.990 & 0.048  \\
    BAO Amp. & 1.006 & 0.016 & 1.008 & 0.055 \\
    Match fitting range & $1.004$ & 0.015 & 1.009 & 0.064 \\
    DESI priors on $\Sigma_{\parallel,\perp,\textrm{fog}}$, & \multirow{2}{*}{1.003} & \multirow{2}{*}{0.014} & \multirow{2}{*}{1.004} & \multirow{2}{*}{0.051} \\
    $\alpha_{\textrm{BB}} = 1$, fix $\gamma_b$, $b_\partial = 0$   \\
    Sample in $\alpha_{\textrm{iso,AP}}$ & 1.000 & 0.014 & 1.001 & 0.050 \\
    Parameterize bias with $b_1$ and $f$ & 0.999 & 0.014 & 1.004 & 0.048 \\
    Rescale $C(k, \mu)$ by $\alpha$ & 0.999 & 0.014 & 0.999 & 0.048

    \end{tabular} 
    \caption{Detailed comparison of our results and the default DESI results for LRG $0.4 < z < 0.6$. The first three rows reproduce the three columns of Table~\ref{tab:abacus_post_recon_tests_appendix}. 
    \label{tab:detailed_comparison_alphas}}
\end{table}

\section{Author affiliations}

\input{affiliation}
\end{document}

%% file: affiliation.tex
\textsuperscript{4}Lawrence Berkeley National Laboratory, 1 Cyclotron Road, Berkeley, CA 94720, USA \\
\textsuperscript{5}Department of Physics, Boston University, 590 Commonwealth Avenue, Boston, MA 02215 USA \\
\textsuperscript{6}Dipartimento di Fisica ``Aldo Pontremoli'', Universit\`a degli Studi di Milano, Via Celoria 16, I-20133 Milano, Italy \\
\textsuperscript{7}INAF-Osservatorio Astronomico di Brera, Via Brera 28, 20122 Milano, Italy \\
\textsuperscript{8}Department of Physics \& Astronomy, University College London, Gower Street, London, WC1E 6BT, UK \\
\textsuperscript{9}Institute of Cosmology and Gravitation, University of Portsmouth, Dennis Sciama Building, Portsmouth, PO1 3FX, UK \\
\textsuperscript{10}Institute for Computational Cosmology, Department of Physics, Durham University, South Road, Durham DH1 3LE, UK \\
\textsuperscript{11}Instituto de F\'{\i}sica, Universidad Nacional Aut\'{o}noma de M\'{e}xico,  Circuito de la Investigaci\'{o}n Cient\'{\i}fica, Ciudad Universitaria, Cd. de M\'{e}xico  C.~P.~04510,  M\'{e}xico  \\
\textsuperscript{12}Department of Astronomy, San Diego State University, 5500 Campanile Drive, San Diego, CA 92182, USA \\
\textsuperscript{13}NSF NOIRLab, 950 N. Cherry Ave., Tucson, AZ 85719, USA \\
\textsuperscript{14}Space Sciences Laboratory, University of California, Berkeley, 7 Gauss Way, Berkeley, CA  94720, USA \\
\textsuperscript{15}University of California, Berkeley, 110 Sproul Hall \#5800 Berkeley, CA 94720, USA \\
\textsuperscript{16}Institut de F\'{i}sica d’Altes Energies (IFAE), The Barcelona Institute of Science and Technology, Edifici Cn, Campus UAB, 08193, Bellaterra (Barcelona), Spain \\
\textsuperscript{17}Departamento de F\'isica, Universidad de los Andes, Cra. 1 No. 18A-10, Edificio Ip, CP 111711, Bogot\'a, Colombia \\
\textsuperscript{18}Observatorio Astron\'omico, Universidad de los Andes, Cra. 1 No. 18A-10, Edificio H, CP 111711 Bogot\'a, Colombia \\
\textsuperscript{19}Institut d'Estudis Espacials de Catalunya (IEEC), c/ Esteve Terradas 1, Edifici RDIT, Campus PMT-UPC, 08860 Castelldefels, Spain \\
\textsuperscript{20}Institute of Space Sciences, ICE-CSIC, Campus UAB, Carrer de Can Magrans s/n, 08913 Bellaterra, Barcelona, Spain \\
\textsuperscript{21}University of Virginia, Department of Astronomy, Charlottesville, VA 22904, USA \\
\textsuperscript{22}Fermi National Accelerator Laboratory, PO Box 500, Batavia, IL 60510, USA \\
\textsuperscript{23}Institut d'Astrophysique de Paris. 98 bis boulevard Arago. 75014 Paris, France \\
\textsuperscript{24}IRFU, CEA, Universit\'{e} Paris-Saclay, F-91191 Gif-sur-Yvette, France \\
\textsuperscript{25}Center for Cosmology and AstroParticle Physics, The Ohio State University, 191 West Woodruff Avenue, Columbus, OH 43210, USA \\
\textsuperscript{26}Department of Physics, The Ohio State University, 191 West Woodruff Avenue, Columbus, OH 43210, USA \\
\textsuperscript{27}The Ohio State University, Columbus, 43210 OH, USA \\
\textsuperscript{28}Department of Physics, University of Michigan, 450 Church Street, Ann Arbor, MI 48109, USA \\
\textsuperscript{29}University of Michigan, 500 S. State Street, Ann Arbor, MI 48109, USA \\
\textsuperscript{30}Department of Physics, The University of Texas at Dallas, 800 W. Campbell Rd., Richardson, TX 75080, USA \\
\textsuperscript{31}Department of Physics, Southern Methodist University, 3215 Daniel Avenue, Dallas, TX 75275, USA \\
\textsuperscript{32}Department of Physics and Astronomy, University of California, Irvine, 92697, USA \\
\textsuperscript{33}Sorbonne Universit\'{e}, CNRS/IN2P3, Laboratoire de Physique Nucl\'{e}aire et de Hautes Energies (LPNHE), FR-75005 Paris, France \\
\textsuperscript{34}Departament de F\'{i}sica, Serra H\'{u}nter, Universitat Aut\`{o}noma de Barcelona, 08193 Bellaterra (Barcelona), Spain \\
\textsuperscript{35}Instituci\'{o} Catalana de Recerca i Estudis Avan\c{c}ats, Passeig de Llu\'{\i}s Companys, 23, 08010 Barcelona, Spain \\
\textsuperscript{36}Department of Physics and Astronomy, Siena University, 515 Loudon Road, Loudonville, NY 12211, USA \\
\textsuperscript{37}Departamento de F\'{\i}sica, DCI-Campus Le\'{o}n, Universidad de Guanajuato, Loma del Bosque 103, Le\'{o}n, Guanajuato C.~P.~37150, M\'{e}xico \\
\textsuperscript{38}Instituto Avanzado de Cosmolog\'{\i}a A.~C., San Marcos 11 - Atenas 202. Magdalena Contreras. Ciudad de M\'{e}xico C.~P.~10720, M\'{e}xico \\
\textsuperscript{39}Instituto de Astrof\'{i}sica de Andaluc\'{i}a (CSIC), Glorieta de la Astronom\'{i}a, s/n, E-18008 Granada, Spain \\
\textsuperscript{40}Departament de F\'isica, EEBE, Universitat Polit\`ecnica de Catalunya, c/Eduard Maristany 10, 08930 Barcelona, Spain \\
\textsuperscript{41}Department of Physics and Astronomy, Sejong University, 209 Neungdong-ro, Gwangjin-gu, Seoul 05006, Republic of Korea \\
\textsuperscript{42}Abastumani Astrophysical Observatory, Tbilisi, GE-0179, Georgia \\
\textsuperscript{43}Department of Physics, Kansas State University, 116 Cardwell Hall, Manhattan, KS 66506, USA \\
\textsuperscript{44}Faculty of Natural Sciences and Medicine, Ilia State University, 0194 Tbilisi, Georgia \\
\textsuperscript{45}CIEMAT, Avenida Complutense 40, E-28040 Madrid, Spain \\
\textsuperscript{46}National Astronomical Observatories, Chinese Academy of Sciences, A20 Datun Road, Chaoyang District, Beijing, 100101, P.~R.~China \\